\documentclass[10pt]{article}
\makeatletter\def\input@path{{./}{arxiv/source/}}\makeatother
\usepackage{amsmath}
\usepackage{amssymb}
\usepackage{amsthm}
\usepackage{mathtools}
\usepackage{multirow}
\usepackage{array}
\usepackage{booktabs}
\usepackage{enumitem}
\usepackage[sc]{mathpazo}
\usepackage[scaled=0.92]{helvet}
\usepackage[T1]{fontenc}
\usepackage{microtype}
\usepackage{graphicx}
\usepackage[letterpaper,margin=0.75in,columnsep=0.25in]{geometry}
\makeatletter
\renewcommand{\fnum@figure}{Figure~\thefigure}
\renewcommand{\fnum@table}{\textbf{Table \thetable}}
\long\def\@makecaption#1#2{\vskip\abovecaptionskip{\small\def\textbf##1{{\sffamily\bfseries ##1}}\textbf{#1}\enspace #2\par}\vskip\belowcaptionskip}
\makeatother
\usepackage[round,semicolon]{natbib}
\setcitestyle{aysep={},yysep={,}}
\usepackage{url}
\allowdisplaybreaks[4]

\newtheorem{theorem}{Theorem}
\newcommand{\eps}{\varepsilon}
\newcommand{\SI}{\emph{SI Appendix}}
\newcommand{\addtabletext}[1]{\par\smallskip{\footnotesize\leftskip=0pt\rightskip=0pt\parfillskip=0pt plus 1fil\relax #1\par}}

\def\scititle{Competitive exploration in structured environments}
\makeatletter
\renewcommand\section{\@startsection{section}{1}{\z@}{-3ex \@plus -1ex \@minus -.2ex}{1.2ex \@plus .2ex}{\sffamily\bfseries\Large}}
\renewcommand\subsection{\@startsection{subsection}{2}{\z@}{-2.5ex \@plus -1ex \@minus -.2ex}{1ex \@plus .2ex}{\sffamily\bfseries\large}}
\renewcommand\subsubsection{\@startsection{subsubsection}{3}{\z@}{-2ex \@plus -1ex \@minus -.2ex}{0.6ex \@plus .2ex}{\sffamily\bfseries\itshape\normalsize}}
\renewcommand\paragraph{\@startsection{paragraph}{4}{\z@}{1.5ex \@plus .5ex \@minus .2ex}{-1em}{\sffamily\bfseries\small}}
\makeatother
\usepackage{xcolor}
\usepackage{etoolbox}
\AtBeginEnvironment{table}{\setlength{\belowcaptionskip}{4pt}}
\usepackage[colorlinks=true,allcolors={rgb:red,1;green,3;blue,8}]{hyperref}

\begin{document}
\twocolumn[{%
\begin{flushright}
{\sffamily\bfseries\huge \scititle\par}
\vspace{12pt}
{\sffamily\bfseries\small James P.~Wolfe$^{1}$, Joshua B.~Plotkin$^{2,\ast}$\par}
\vspace{3pt}
{\sffamily\scriptsize $^{1}$Graduate Group in Applied Mathematics and Computational Science, University of Pennsylvania, Philadelphia, PA 19104, USA, $^{2}$Department of Biology, University of Pennsylvania, Philadelphia, PA 19104, USA\par
$^\ast$Corresponding author. Email: jplotkin@sas.upenn.edu\par
\vspace{3pt}
October 25, 2026\par}
\end{flushright}
\vspace{22pt}
{\sffamily\small\noindent\textbf{ABSTRACT}\enspace Organisms must constantly weigh exploiting the resources they know against exploring for better ones. When foragers compete, the best behavior for each forager also depends on what its competitors do. How does competition change the strategic balance between exploration and exploitation? Here we address this question in a minimal model: two $\eps$-greedy temporal-difference learners on a one-dimensional lattice, who start at a small reward at one end and are unaware of a larger reward at the other. For a lone forager, the optimal exploration rate increases monotonically with the value of the distant reward, and with its distance up to a critical distance. When two foragers share the lattice and split rewards, the choice of exploration rate becomes a game whose Nash equilibria have both foragers exploring, both staying, or one exploring while the other stays (anti-coordination).
Under local best-response dynamics, pairs of foragers often converge on both exploring or both staying, even when anti-coordination would be socially optimal. This social dilemma arises for a wide range of parameters, and it persists even in a simplified model without explicit spatial structure. A minimal, naturalistic signal, in which one forager informs the other that it has found the distant reward, can remove many of the dilemmas in which both foragers explore, and can produce a Nash equilibrium whose collective payoff exceeds that of every outcome available without signaling.\par}
\vspace{24pt}
}]

Across species and timescales, organisms must balance exploiting the resources they already know against exploring for better ones they have not yet found. A bacterium performing chemotaxis alternates between swimming up an attractant gradient and tumbling to reorient at random \citep{berg1972-ChemotaxisEscherichiaColi}; a bee colony works its known flower patches while dispatching scouts in search of richer ones; and a person searching for information faces the same choice \citep{pirolli1999-InformationForaging}. The tradeoff has been studied for decades in behavioral ecology, where optimal foraging theory asks how an animal should time its departure from a depleting resource and divide its effort among sites of differing quality \citep{charnov1976-OptimalForagingMarginal,stephens1986-ForagingTheory,macarthur1966-OptimalUsePatchy}. Whether the relevant timescale is a single lifetime or an evolutionary one, the balance that maximizes long-run reward is difficult to strike, and it depends both on the structure of the environment and on the behavior of others exploiting it.

Reinforcement learning provides a formal framework for managing the exploration--exploitation tradeoff \citep{sutton2018-ReinforcementLearningIntroduction}. An agent that interacts repeatedly with its environment maintains value estimates for the states and actions available to it, and at each step chooses between the action favored by its current value estimates and an alternative that may prove better. The tension is sharpest in the multi-armed bandit problem \citep{robbins1952-AspectsSequentialDesign}, in which an agent chooses repeatedly among fixed options with unknown reward distributions. Most of the resulting theory pursues regret minimization---algorithms whose cumulative reward provably stays close to that of the best fixed option \citep{lai1985-AsymptoticallyEfficientAdaptive,auer2002-FinitetimeAnalysisMultiarmed}. These guarantees are worst-case: they bound performance across all environments rather than prescribe how much an agent should explore in a particular one.
How humans and other animals in fact resolve the exploration--exploitation tradeoff has itself become an active question in cognitive science and neuroscience \citep{daw2005-UncertaintybasedCompetitionPrefrontal,tavoni2022-HumanInferenceReflectsa}.

Animal foraging departs from the stylized bandit problem in one important respect: foraging options are not equally accessible. Reaching a distant resource takes time, during which the forager collects nothing, so exploration incurs an opportunity cost in travel time that a bandit model omits. We therefore study reinforcement learners in spatially structured environments, in which movement between options is itself costly. We adopt temporal-difference (TD) learning \citep{sutton1988-LearningPredictMethods}, in which an agent revises the value it assigns to a site using the values it has assigned to neighboring sites. TD learning is a standard model of animal reward learning, and it has a well-documented neural correlate: the phasic activity of midbrain dopamine neurons covaries with the reward-prediction error term of the TD update \citep{schultz1997-NeuralSubstratePrediction,montague1996-FrameworkMesencephalicDopamine}, and the same signal has been identified in the human brain \citep{odoherty2003-TemporalDifferenceModels}. TD learning has been applied specifically to spatial foraging, such as bumblebees moving among flowers \citep{montague1995-BeeForagingUncertain}.

How much should a learner explore new territory versus exploit known resources? The simplest rule is $\eps$-greedy, governed by a single exploration rate $\eps$: with probability $1-\eps$ the agent takes the action it currently values most, and with probability $\eps$ it moves at random. Despite its simplicity, $\eps$-greedy often matches or outperforms more elaborate algorithms with dynamic exploration rates and  stronger convergence guarantees \citep{vermorel2005-MultiarmedBanditAlgorithms,kuleshov2014-AlgorithmsMultiarmedBandit}, and it requires no sophisticated cognition about the environment, unlike dynamic programming
\citep{bellman1957-DynamicProgramming,puterman1994-MarkovDecisionProcesses,sutton2018-ReinforcementLearningIntroduction}. But systematic guidance on which exploration rate $\eps$ performs best in a given environment remains scarce. A fixed rate is also the natural object of study for an evolved forager: an animal may be unable to adjust its exploration rate from moment to moment, and instead have an intrinsic rate shaped by natural selection to suit its ecology \citep{dingemanse2010-RecentModelsAdaptive,wolf2007-LifehistoryTradeoffsFavour}. Consistent with such intrinsic  variation, ant colonies of different species differ in their exploitation intensity: species that commit most of their foragers to the better of two feeders are slow to reallocate when the poorer feeder improves, whereas species that spread foragers more evenly shift readily to the improved feeder \citep{latty2013-KeepingTrackChanges}. In the terms of our model, each species behaves as though it has a different fixed exploration rate.

Foragers are rarely alone, though, and competition changes what each of them should do \citep{giraldeau2000-SocialForagingTheory}. Competitive multi-armed bandit models with shareable rewards address a version of this question: at the Nash equilibrium of such a game the number of agents at each option is roughly proportional to that option's value \citep{xu2023-CompetingShareableArms}, recalling the ideal free distribution of classical foraging theory \citep{fretwell1969-TerritorialBehaviorOther,parker1986-IdealFreeDistributions}. Our setting is neither ideal nor free: the agents lack perfect information, and although nothing prevents a forager from traveling to the distant reward, the opportunity cost of doing so can be large.

Here we ask what fixed exploration rates are individually and socially optimal when foragers compete for shared rewards in a spatially structured environment: a one-dimensional lattice with a small, known reward at one end and a larger, undiscovered reward at the other. We first find that, for a single forager, the reward-maximizing exploration rate is intermediate---rising with the value of the distant reward, and rising and then falling with the distance to it. We then find that competition makes the choice of rate a two-player game whose Nash equilibria include anti-coordination, in which two identical foragers adopt opposite behaviors, and that the Nash equilibria at which both foragers stay or both foragers explore are often collectively worse than anti-coordination, producing a social dilemma. Finally, we find that a simple, cognitively undemanding signal---one forager informing another that it has found the distant reward---can remove the suboptimal equilibrium at which both foragers explore and resolve the dilemma.

\section*{Model}

\subsection*{A spatial model of exploration and exploitation}

\begin{figure}[tbhp]
\centering
\includegraphics[width=\linewidth]{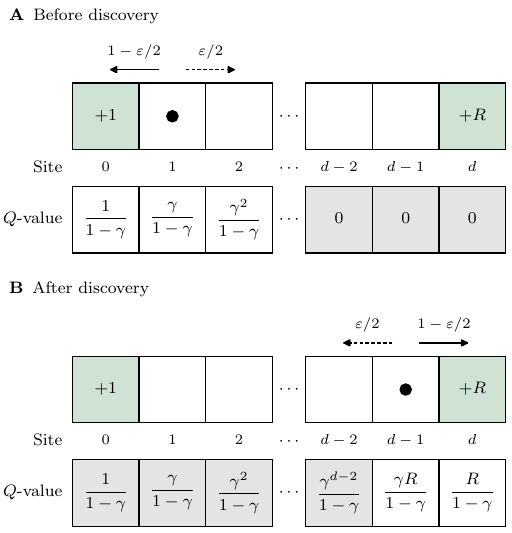}
\caption{\textbf{The biased random walk and the temporal-difference learner describe the same movement rule.} (A) Before it has discovered the distant site $d$, an agent in the interior of the lattice steps toward the near reward at site $0$ with probability $1-\eps/2$ and away from it with probability $\eps/2$. The reward at site $0$ is known to the agent, the interior sites have no reward, and the reward at the far end of the lattice is unknown until the agent first reaches it. A shaded $Q$-value indicates a site that the agent has not yet visited, and whose value it has therefore not updated. (B) An agent that has discovered $d$ is now biased towards it. $Q$-values to the right are now higher, even if the agent has not yet updated the $Q$-values to the left. Shaded $Q$-values now indicate that the agent has not visited these sites since discovering the reward at $d$.}
\label{fig:cartoon}
\end{figure}

Our model represents the essential tension of a foraging task: at each moment an agent must decide whether to exploit a resource it already knows or to explore in search of a better one it has not yet found. We situate this decision in a spatially structured environment, so that exploration incurs an opportunity cost---time spent traveling away from a known reward (Fig.~\ref{fig:cartoon}). An agent begins at a small, familiar reward, unaware of a larger reward far away. A single parameter, the exploration rate $\eps$, governs how readily it explores. Below we specify the environment, the movement policy, and the policy's equivalent descriptions as a temporal-difference learner or as a Markov chain.

We represent the environment as a finite, one-dimensional lattice of sites $\{0, 1, 2, \dots, d\}$, with $d>0$. Two sites contain resources: a small reward at one boundary, $r(0) = 1$, and a larger reward at the other boundary, $r(d) = R > 1$; all sites in between are empty, $r(s) = 0$. The lattice length $d$ thus sets the distance an agent must travel to reach the richer resource, and the reward ratio $R$ governs the benefit of discovering the richer reward. At each discrete time step the agent occupies a single site $s$ and accrues its reward $r(s)$; rewards are non-depleting, so an agent collects $r(s)$ on every step it remains at site $s$. Movement is local: from site $s$ the agent may step only to a site in its neighborhood $n(s)$ (self-inclusive), e.g.\ $n(2) = \{1,2,3\}$ for $d > 2$ and $n(0) = \{0, 1\}$.

The agent moves on this lattice by an $\eps$-greedy rule. At each step it chooses between two behaviors:
\begin{itemize}[leftmargin=*,itemsep=0pt,parsep=0pt,topsep=3pt]
    \item with probability $1-\eps$ it \emph{exploits}, moving toward the highest reward it has so far visited;
    \item with probability $\eps$ it \emph{explores}, moving to a randomly chosen neighboring site (excluding its current one).
\end{itemize}
This rule generates a two-phase biased random walk. Before the agent has discovered the far reward, its only known resource is the near one at site $s=0$, so exploitation biases it leftward: in the interior $\{1, \dots, d-1\}$ it steps left with probability $1-\eps/2$ and right with probability $\eps/2$. Once it has visited site $d$, its bias reverses---it steps right with probability $1-\eps/2$ and left with probability $\eps/2$---and it is biased toward the richer resource. The exploration rate $\eps$ therefore sets the strength of the bias in each phase: an agent that never explores ($\eps = 0$) remains at the near reward indefinitely, whereas larger $\eps$ increases the variance of the walk and the probability of reaching site $d$.

The two-phase biased random walk is what a temporal-difference (TD) learner would produce, as we make precise below. The intuition is that, once the agent reaches the far reward, subsequent steps propagate its value backward along the lattice: each visited site's value estimate is set to a discounted fraction of $R$, decreasing geometrically with distance from site $d$. For a sufficiently large discount factor $\gamma$, the value estimates of the sites visited since that discovery therefore increase monotonically toward site $d$, so that the exploitation move at every interior site points toward it.

The two ends of the lattice require rules of their own. Before site $d$ has been found, site $0$ is reflecting: at site $s=0$ the agent stays with probability $1-\eps$ and steps to site $s=1$ with probability $\eps$. After site $d$ has been found, site $d$ is reflecting in the same way: the agent stays there with probability $1-\eps$ and steps to $d-1$ with probability $\eps$. After site $d$ has been found, an agent at site $s=0$ moves to site $s=1$ with probability $1$, consistent with the interior rule: with probability $1-\eps$ it exploits by moving toward $d$, and with probability $\eps$ it explores to its only available neighbor, site $1$.

An agent values immediate rewards more than future ones, which we encode with a discount factor $\gamma \in (0,1)$: a reward collected $t$ steps in the future is worth a fraction $\gamma^t$ of its immediate value. Discounting represents a finite foraging horizon, and corresponds to averaging reward over a characteristic time $T \approx 1/(1-\gamma)$ (see \emph{Markov chain formulation} below). Discounting also determines which reward an agent prefers once it has found both, since a more distant reward takes more time to reach and is therefore discounted more heavily. From an interior site $s \in \{1, \dots, d-1\}$, exploiting toward site $0$---a journey of $s$ steps followed by an indefinite stay---yields expected discounted reward
\begin{multline*}r(s) + \gamma r(s-1) + \gamma^2 r(s-2) + \dots \\ + \left(\gamma^s + \gamma^{s+1} + \dots\right) r(0) = \frac{\gamma^s}{1-\gamma},\end{multline*}
using $r(0) = 1$ and $r(s) = 0$ for $0 < s < d$. Exploiting toward site $d$ instead---a journey of $d-s$ steps---yields $\gamma^{d-s}R/(1-\gamma)$. The exploitation move is therefore toward whichever reward has the larger discounted value: toward site $0$ when $\gamma^s > \gamma^{d-s}R$, and toward site $d$ when $\gamma^s < \gamma^{d-s}R$. With discounting, the policy for an agent at site $s$ becomes:
\begin{itemize}[leftmargin=*,itemsep=0pt,parsep=0pt,topsep=3pt]
    \item With probability $1-\eps$ it \emph{exploits}, moving toward (or staying at) site $0$ if it has not seen reward $R$ or if $\gamma^s > \gamma^{d-s}R$, and toward (or staying at) site $d$ if it has seen reward $R$ and $\gamma^s < \gamma^{d-s}R$.
    \item With probability $\eps$ it \emph{explores}, moving to a randomly chosen neighboring site (excluding its current one).
\end{itemize}
In the boundary case $\gamma^s = \gamma^{d-s}R$ the agent moves left or right with equal probability. We work in the regime of large $\gamma$, where $\gamma^s < \gamma^{d-s}R$ for every site including $s=0$, equivalently $\gamma^d R>1$, so that any agent that has discovered site $d$ is biased toward that side from everywhere on the lattice. In this \emph{high-discount regime}, which we assume throughout unless noted otherwise, the discounted policy reduces to the two-phase biased random walk described above, and the movement rule no longer depends on $\gamma$.

\subsection*{Interpretation as a temporal-difference or $Q$-learning problem}
The movement rules outlined above also arise from temporal-difference learning \citep{sutton1988-LearningPredictMethods,sutton2018-ReinforcementLearningIntroduction}, which is a standard model for how animals and algorithms learn to value their environment. Suppose the agent maintains an estimate $Q(s)$ of the long-term value of each site $s$---the total discounted reward it expects starting from $s$---initialized at $Q(s) = 0$ everywhere. Upon first visiting a site $s$ that provides a reward $r(s) > 0$, the agent assesses its steady-state discounted value $r(s)/(1-\gamma)$ and sets $Q(s) = r(s)/(1-\gamma)$; that is, on encountering a reward the agent immediately registers its long-run worth. Updating $Q$-values to account for the discount factor is a modeling choice made without loss of generality: an alternative assignment based on instantaneous rewards, $Q(s) = r(s)$, scales every $Q$-value by $1-\gamma$ and so leaves every comparison between sites, and with it the movement rules, unchanged (cf.~\citep{wiewiora2003-PotentialBasedShapingQValue}).

Because the agent begins at site $0$, the rule for assigning $Q$-values fixes $Q(0) = 1/(1-\gamma)$ immediately, while all other sites are initialized with $Q=0$. The far site keeps $Q(d) = 0$ until the agent first reaches it: the value $Q(d) = R/(1-\gamma)$ is assigned on reaching site $d$, not in advance. The agent thus has no knowledge of the far reward until exploration reveals it, at which time the agent computes its long-term discounted value.

At the empty sites between the two rewards ($s \in \{1,2,\ldots, d-1\}$ where $r(s)=0$), the agent updates its estimate by temporal-difference learning, which we call state-value $Q$-learning:
\[
Q(s) = r(s) + \gamma \max\{Q(s') : s' \in n(s)\}.
\]
This recursion---a site's value equals its immediate reward plus the discounted value of its best neighbor---is the standard Bellman update used in $Q$-learning. Two choices adapt it to our setting. First, because the reward at each site is fixed rather than noisy, the agent overwrites its old estimate outright rather than averaging old and new; in the language of $Q$-learning, the learning rate is $\alpha = 1$. Second, the agent stores a single value per site rather than a separate value for each available move, because a site's location already determines which moves are possible. The agent updates only the value of the site it occupies, so value estimates propagate one site at a time as the walk proceeds.

Applying the Bellman update repeatedly as the agent moves produces $Q$ values that are monotone across the sites visited in each phase of the walk. Before the agent has visited site $d$, each site it has visited has value $Q(s) = \gamma^s/(1-\gamma)$: the first step sets $Q(1) = \gamma/(1-\gamma)$ from the reward at site $0$, the next sets $Q(2) = \gamma^2/(1-\gamma)$, and so on. Sites not yet visited---including site $d$---remain at their initial value $Q=0$. Once the agent has reached site $d$, a similar sequence of updates proceeds inward from the far reward: in the high-discount regime each subsequently visited site is updated to $Q(s) = \gamma^{d-s}R/(1-\gamma)$, so that walks are biased towards the distant reward.

More generally, for an arbitrary discount factor, the $Q$ values of sites visited after discovery of $d$ will satisfy
\[
Q(s) = \max\left\{\frac{\gamma^s}{1-\gamma}, \frac{\gamma^{d-s}R}{1-\gamma}\right\},
\]
which means that post-discovery walks are biased to the left closer to $s=0$ and to the right closer to $s=d$. In general, moving to the neighboring site with highest $Q$ value therefore moves the agent in the direction that maximizes its expected discounted reward, given its current knowledge of the reward landscape. In the high-discount regime, where an agent that has found site $d$ is right-biased at every interior site, the TD formulation coincides exactly with the two-phase biased random walk described above.

\subsection*{Markov chain formulation} \label{sec:transition-matrix}

The movement of the agent, even outside the high-discount regime, can be reduced to a Markov chain on $2(d+1)$ states. The Markov chain describes the same movement as the TD learner, but it gives the agent's expected reward exactly, without simulating many runs. At $d=8$ and $R=5$, the Markov chain prediction matches the mean reward of $5000$ TD-learner runs (\SI, Fig.~\ref{sup-fig:markov_vs_q}).

Enumerating the states reachable from each state under the movement rules above gives a transition matrix $P$ with at most two nonzero entries per row (see \emph{Materials and Methods}). We write each state as a pair $(s, B)$, where $s \in \{0, 1, \dots, d\}$ is the agent's current site and $B \in \{0,1\}$ flags whether or not it has yet visited site $d$. Let $\mathbf{p}_0$ be the initial distribution over states---here concentrated on state $(0,0)$---and let $\mathbf{r}$ be the reward vector, with $\mathbf{r} = 1$ on states $(0,B)$, $\mathbf{r} = R$ on state $(d,1)$, and $0$ elsewhere. The expected cumulative reward over $T+1$ steps is $\mathbf{p}_0\sum_{t=0}^T P^t \mathbf{r}$. We instead use the discounted sum over an infinite horizon, with the agent's own discount factor $\gamma$, which admits a closed form:
\[\pi = \mathbf{p}_0\sum_{t=0}^\infty \gamma^t P^t \mathbf{r} = \mathbf{p}_0(I-\gamma P)^{-1}\mathbf{r}.\]
Setting $\gamma = 1-1/T$ recovers results close to---though not identical to---the finite-horizon sum, consistent with the characteristic time $T \approx 1/(1-\gamma)$ noted above.

\subsection*{Two competing agents}

We now add a second, identical agent to the same environment. When both agents occupy the same site at the same time, they split its reward. We assume $R > 2$, so that an agent's share of the far reward, $R/2$, still exceeds the near reward $r(0)=1$, which is necessary for an agent that has discovered site $d$ to stay biased towards it even when it must share.

Under the TD dynamics, an agent that has found site $d$ remains biased toward it from every site whenever $\gamma^d R/2 > 1$. (On reaching $d$, an agent sets $Q(d) = R/(1-\gamma)$ when foraging alone and $Q(d) = R/[2(1-\gamma)]$ when sharing the reward.) We work in this regime, choosing $R$, $d$, and $\gamma$ so that $\gamma^d R/2 > 1$. Each agent's reward then depends on the other's location, but its local movement rule does not, so we can reuse the single-agent transition matrix and adjust only the reward vector to account for splitting.

The parameters of the model, their interpretations, and the ranges we consider are collected in Table~\ref{tab:parameters}.

\begin{table}[tbhp]
\centering
\footnotesize
\caption{Parameters of the model}\label{tab:parameters}
\begin{tabular}{@{}c>{\raggedright\arraybackslash\hyphenpenalty=10000}p{4.4cm}>{\raggedright\arraybackslash\hyphenpenalty=10000}p{2.3cm}@{}}
\toprule
Symbol & Interpretation & Range \\
\midrule
$d$ & distance to the distant reward (lattice length) & integer $>0$ \\
$R$ & reward ratio, distant reward relative to the near reward & $R>1$ ($R>2$ with two agents) \\
$\gamma$ & discount factor, setting the foraging horizon $T\approx 1/(1-\gamma)$ & $0<\gamma<1$ \\
$\eps$ & exploration rate, an agent's strategy & $0\le\eps\le1$ \\
$\eps'$ & exploration rate adopted after a competitor's signal that it has reached the distant reward (signaling model) & $0\le\eps'\le1$ \\
\bottomrule
\end{tabular}
\end{table}

\section*{Results}

\begin{figure*}[tb]
\centering
\includegraphics[width=\linewidth]{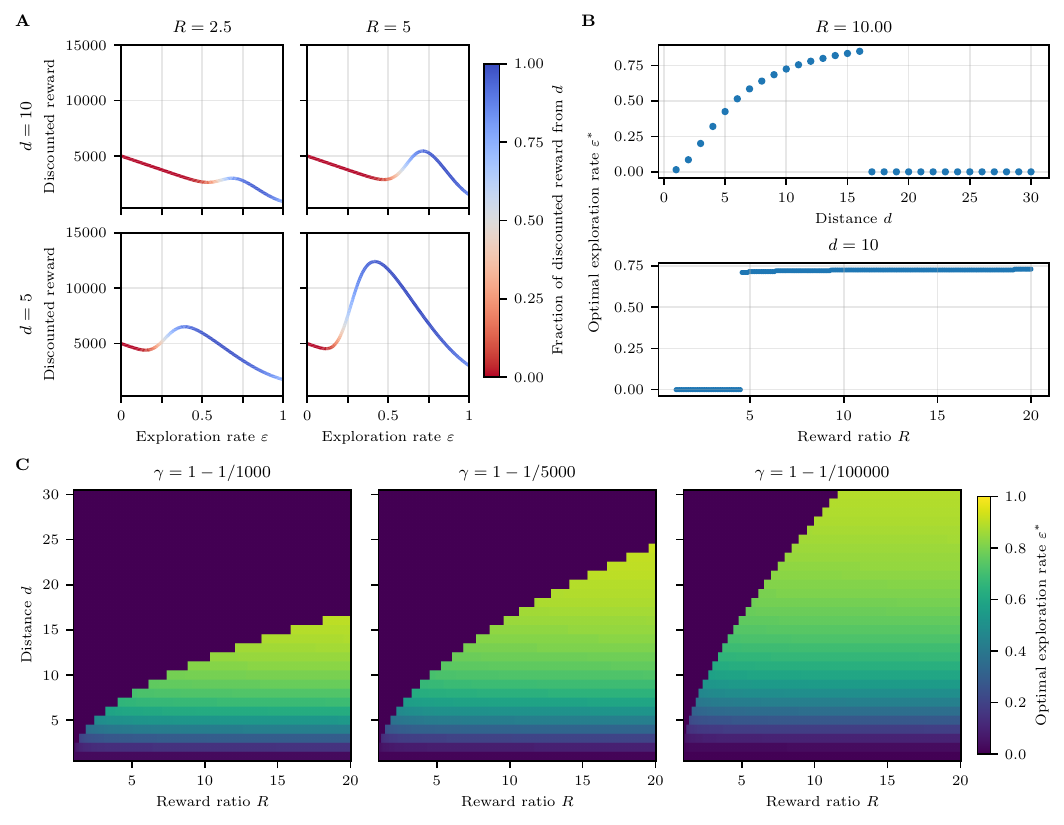}
\caption{\textbf{An intermediate exploration rate is optimal for distant, valuable rewards, and the optimum is non-monotonic in the distance.} (A) Expected discounted reward $\pi(\eps)$ of a single agent as a function of its exploration rate $\eps\in[0,1]$, for four combinations of distance and reward: rows give the distance ($d=10$ top, $d=5$ bottom) and columns the reward ratio ($R=2.5$ left, $R=5$ right). Each curve has a local maximum at $\eps=0$ and, except in extreme regimes, a second local maximum at an intermediate rate. The color of each curve gives the fraction of the discounted reward that the agent collects from the distant site, from none in red to all of it in blue. (B) The reward-maximizing rate $\eps^*$ as the distance $d$ varies (top; reward ratio fixed at $R=10$) and as the reward ratio $R$ varies (bottom; distance fixed at $d=10$): the optimal rate rises with the distance and then falls discontinuously to zero, and jumps from zero to a positive value as the reward ratio crosses a threshold. (C) The optimal rate $\eps^*$ over the plane of reward ratio $R$ and distance $d$, for three discount factors $\gamma$, from $\eps^*=0$ in dark to $\eps^*=1$ in light. A sharp boundary separates the region of pure exploitation ($\eps^*=0$) from the region in which some exploration is optimal ($\eps^*>0$), and the exploring region widens as the foraging horizon lengthens. Parameters: $R=1.05$--$20$, $d=1$--$30$, $\gamma\in\{1-1/1000,\,1-1/5000,\,1-1/100000\}$ in (C); $\gamma=1-1/5000$ in (A) and (B).}
\label{fig:single-agent-optimal}
\end{figure*}

\subsection*{An intermediate exploration rate is optimal for distant, valuable rewards} \label{sec:optimal-exploration-rate}

What fixed exploration rate $\eps$ maximizes a single agent's expected discounted reward, $\pi(\eps)$? Except in extreme regimes, the expected discounted reward is non-monotonic in the exploration rate, with one local maximum at $\eps = 0$ and another
at an intermediate exploration rate $\eps > 0$ (Fig.~\ref{fig:single-agent-optimal}A). Which of the two exploration rates is the global reward maximizer, $\eps^*$, depends on the relative value of the distant reward, $R$, and the distance to reach it, $d$.

Pure exploitation ($\eps^* = 0$) is optimal when the distant reward is small or far away (e.g.\ $R = 2.5$, $d = 10$): the opportunity cost of travel outweighs
the gain from reaching site $d$. Increasing the value of the distant reward ($R = 5$, $d = 10$) or
shortening the distance ($R = 2.5$, $d = 5$) moves the optimal exploration rate to the interior, so that some exploration is optimal. As the distance $d$ grows, this interior optimum
shifts to larger values of $\eps$, and less of the agent's total reward is collected from the distant site $s=d$.

The intuition is a tension between finding the distant reward and
exploiting it. Some exploration is needed to reach site $d$ at all, and the
farther it is, the more exploration this requires.
But the exploration rate is fixed, so an agent that explores enough to reach site $s=d$ also steps away from it at that same rate, forfeiting reward in the process. The optimal exploration rate balances these two effects; when the reward is too small or too distant for the gain from reaching $d$ to outweigh the
loss due to continued exploration, the optimum is $\eps^* = 0$.

A sharp boundary separates the parameters at which exploiting is optimal from those at which exploring is optimal (Fig.~\ref{fig:single-agent-optimal}B,C). For fixed distance $d$, the optimum stays
at $\eps^* = 0$ until the reward benefit $R$ crosses a critical value $R^*(d)$, and then it jumps to a
positive rate and increases gradually. For fixed benefit $R$, the optimal exploration rate increases with $d$ up to
a critical distance $d^*(R)$, and then it drops discontinuously to $\eps^* = 0$. The
critical reward ratio $R^*(d)$ increases with the distance $d$, faster than linearly beyond the shortest distances, and it decreases as the foraging horizon lengthens (\SI, Fig.~\ref{fig:threshold}).

Away from the boundary between exploiting and exploring, the optimal exploration rate is robust to random variation in the distance to the distant reward. When the distance is random, $d \sim \mathrm{Poisson}(\lambda) + 1$ with mean distance $\lambda+1$, the optimal exploration rate closely matches that of the static environment with $d = \lambda + 1$; near that boundary the two optimal rates differ, in either direction depending on the discount factor
(see \SI, Fig.~\ref{sup-fig:poisson_opt_eps}). Hence a single exploration rate, chosen for the mean distance, is close to the best fixed rate for a distance that varies about that mean, except near the boundary between exploiting and exploring.

\begin{figure*}[tb]
\centering
\includegraphics[width=\linewidth]{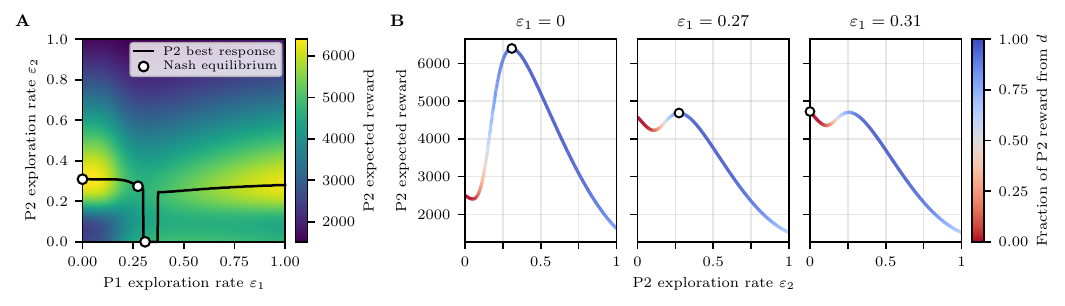}
\caption{\textbf{A competitor's best response is non-monotonic in the focal agent's exploration rate, and at a short distance and a small reward ratio the game can have three Nash equilibria.} (A) Player 2's expected discounted payoff across a fine grid of exploration-rate pairs $(\eps_1,\eps_2)$, from low in dark to high in light. The black curve is player 2's best response---the rate that maximizes its payoff given player 1's rate---and it is non-monotonic in $\eps_1$. White circles mark the three Nash equilibria, where the best-response curve meets its reflection across the diagonal: one symmetric pair at which both agents explore, and a pair of anti-coordination equilibria related by exchanging the players' labels. (B) Slices of player 2's payoff at three fixed values of player 1's rate, $\eps_1 = 0$, $0.27$, and $0.31$, with the white circle marking player 2's best response in each. Player 2 faces a reward curve of a similar shape to the single-agent reward curve, and changing player 1's rate deforms that curve in the way that changing the reward ratio $R$ deforms the single-agent one, exactly so when player 1 never explores. Between $\eps_1=0.27$ and $\eps_1=0.31$ the best response moves discontinuously from the interior maximum to $\eps_2=0$. Curve color gives the fraction of player 2's reward collected from the distant site, from none in red to all of it in blue. Parameters: $d=4$, $R=2.1$, $\gamma=1-1/5000$.}
\label{fig:payoff-mat}
\end{figure*}

\subsection*{Two competing foragers play a game with multiple equilibria}

\subsubsection*{Strategies and payoffs}\label{sec:strategies}

How should an agent explore when it must compete with another agent for the same rewards? Rewards at a shared site are split, and so the payoff to a focal agent with exploration
rate $\eps_1$ depends not only on the distance $d$ to receive a reward benefit $R$, but also on the exploration rate $\eps_2$ of the other agent. The choice of exploration rate therefore defines a
symmetric two-player game with continuous strategies: each agent's strategy
is its exploration rate $\eps \in [0,1]$, its payoff is the expected
discounted reward $\pi(\eps_1, \eps_2; d, R, \gamma)$ it collects, and the distance $d$,
reward benefit $R$, and discount factor $\gamma$ enter as exogenous
parameters. We ask what exploration rates two self-interested agents will tend to
adopt---and whether or not those rates maximize the payoff the two of them collect together.

A competitor's best response is non-monotonic in the focal agent's exploration rate (Fig.~\ref{fig:payoff-mat}A). Because each player's strategy is continuous, the game is described by a payoff function $\pi_1(\eps_1, \eps_2)=\pi_2(\eps_2,\eps_1)$, which becomes a payoff matrix when each player's exploration rate is discretized on a fine grid (Fig.~\ref{fig:payoff-mat}A, for distant reward benefit $R = 2.1$ and distance $d = 4$). Player 2's best response function is the exploration rate that maximizes its payoff given player 1's rate, $\mathrm{BR}(\eps_1) = \mathrm{argmax}_{\eps_2}\,\pi_2(\eps_1, \eps_2)$. When $\eps_1$ is low, player 1 rarely reaches the
distant reward, which is left uncontested, and so player 2's best strategy
is to explore. When $\eps_1$ is moderate, player 1 both finds the distant
reward and exploits it effectively, and so player 2 does best to stay at the
near site $s=0$ by choosing $\eps_2=0$. And when $\eps_1$ is high, player 1 finds the distant
reward but then steps away from it at a high rate, and so exploration is again optimal for player 2.

Against a competitor with a fixed exploration rate, an agent faces a version of the single-agent problem: its reward curve has one local maximum at $\eps_2 = 0$ and another at an intermediate exploration rate $\eps_2 > 0$ (Fig.~\ref{fig:payoff-mat}B). Varying player 1's exploration
rate deforms this curve. The correspondence to the single-agent problem is exact when player 1 never
explores: if $\eps_1 = 0$, player 1 stays at site $s=0$, so player
2 collects only half the near reward whenever it is there, while the distant
reward is uncontested. Player 2 therefore faces an environment equivalent to
the single-agent problem with the distant reward benefit doubled, $R' = 2R$,
and total payoff scaled by a factor of $1/2$.

A competitor that finds and exploits the distant reward lowers the value of that reward to the other agent, since the two players split it whenever both are there. And just as a single agent's optimal
exploration rate drops discontinuously to zero when the reward benefit falls
below the critical value $R^*(d)$, player 2's best response drops
discontinuously---from an intermediate rate to $\eps_2 = 0$---as $\eps_1$
increases. The discontinuity occurs at the value of player~1's exploration rate $\eps_1$ at which the interior local maximum of player~2's reward curve falls below its value at zero exploration, $\eps_2 = 0$, so that the global reward maximizer changes from an intermediate rate to $\eps_2=0$ (Fig.~\ref{fig:payoff-mat}B). Hence player~2's best response has the same structure as the single-agent optimum---an interior exploration rate that drops discontinuously to zero---with the competitor's exploration rate $\eps_1$, rather than the reward ratio $R$, as the quantity that crosses the threshold.

\subsubsection*{Nash equilibria}\label{sec:nash-eq}

A Nash equilibrium is a pair of exploration rates at which neither player can increase its payoff by unilaterally changing its own rate, namely an intersection of the two best-response curves, where $\mathrm{BR}(\eps_1) = \eps_2$ and $\mathrm{BR}(\eps_2) = \eps_1$. Because the game is symmetric, player 1's best-response curve is player 2's reflected across the diagonal $\eps_1 = \eps_2$.
For the example parameters of Fig.~\ref{fig:payoff-mat} (distant reward
benefit $R = 2.1$, distance $d = 4$) there are three Nash equilibria: one in
which both players explore at the same intermediate rate
($\eps_1 = \eps_2 > 0$), and a pair related by exchanging the players'
labels, in which one player stays at the near site while the other explores
($\eps_1 = 0$ and $\eps_2 > 0$, or vice versa). We refer to the symmetric equilibrium
as ``both explore'' and to the two asymmetric equilibria as ``anti-coordination,'' since
the players adopt opposite behaviors.

Across a sweep of the exogenous parameters, the set of Nash equilibria takes one of four forms: \{both explore,
anti-coordination\}, \{both explore\}, \{anti-coordination\}, and
\{both stay\} (Fig.~\ref{fig:nash_+_social_optima}, top row). Only the set \{both explore, anti-coordination\} contains Nash equilibria of two types, and this set occurs only in a narrow interval of reward ratios at the border between the anti-coordination and both-explore regions. At distance $d=4$ and discount factor $\gamma = 1-1/5000$, for example, both the symmetric equilibrium and the anti-coordination pairs are Nash equilibria at reward ratio $R = 2.1$ (Fig.~\ref{fig:payoff-mat}), whereas at $R = 2.15$ the anti-coordination pairs are no longer Nash equilibria. In particular, when both agents staying is a Nash equilibrium it is the only one, because staying is then the best response to every exploration rate of the competitor (\SI, section~\ref{sec:bothstay}).

Competition thus produces a qualitatively new outcome: at an
anti-coordination equilibrium, two identical agents adopt different
behaviors---one stays while the other explores. Moreover, the Nash
equilibria are often multiple, so the game's parameters alone do not
determine which behaviors two self-interested agents will adopt. Which
equilibrium a pair of agents reaches---if any---depends on how they adjust
their exploration rates over time, the question we take up next.

\subsubsection*{Locally stable equilibria}

\begin{figure*}[t]
\centering
\includegraphics[width=0.99\linewidth]{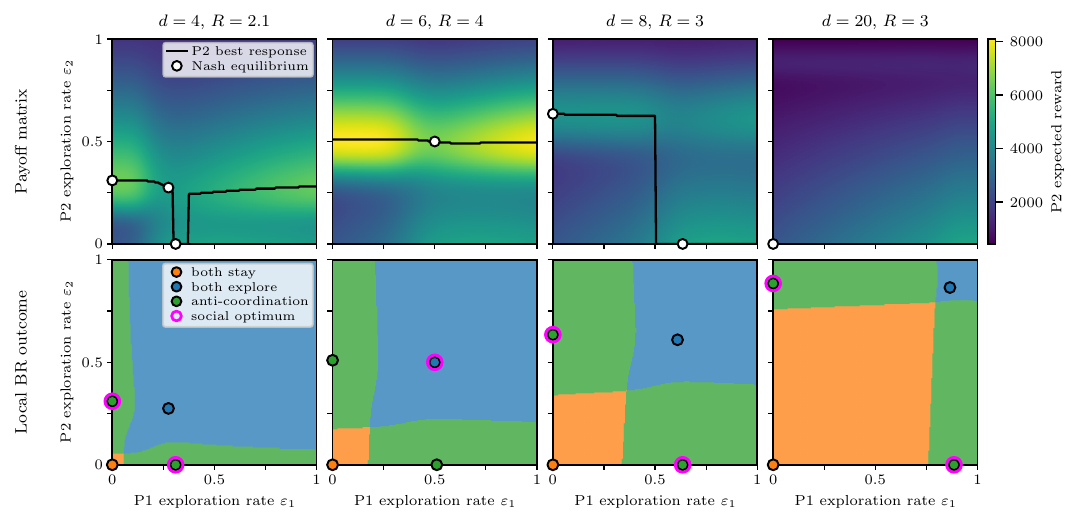}
\caption{\textbf{Each of four parameter regimes has four locally stable equilibria of the same three types, while the basins of attraction from which they are reached change markedly.} Columns give four combinations of distance and reward ratio, $(d,R) = (4,2.1)$, $(6,4)$, $(8,3)$, and $(20,3)$, in order of increasing distance. Top: player 2's expected discounted payoff over the unit square of exploration-rate pairs $(\eps_1,\eps_2)$, from low in dark to high in light, with player 2's best-response curve in black and the Nash equilibria as white circles. Bottom: the outcome reached under the local best-response dynamics from each starting pair, shaded orange where both agents end up staying, blue where both explore, and green at anti-coordination. Circles mark the locally stable equilibria themselves, colored by outcome, and magenta rings mark the locally stable equilibrium or equilibria with the highest collective payoff. Parameters: $\gamma=1-1/5000$.}
\label{fig:local_br_alg}
\end{figure*}

Which strategy pairs do agents reach when they adjust their exploration rates incrementally, and do those pairs coincide with the Nash equilibria? Finding a Nash equilibrium requires each agent to compare its payoff across the entire space of alternative exploration rates, and to perform this computation from the perspective of the opponent as well. Organisms may lack the cognitive capacity for this global comparison; more plausibly, they adjust behaviors incrementally, by small changes tested over the course of many games, or by mutations of small effect across generations. We therefore consider myopically updating agents, who after each game shift their exploration rate by a small amount in the direction that increases their payoff.

After each game, each player either keeps its exploration rate or changes it to the neighboring value just below or just above. Each player
chooses, among these three neighboring rates, written $N(\eps_i)$, the rate that
maximizes its payoff against the other player's current strategy; player 1, for
example, adopts $\mathrm{argmax}_{\eps \in N(\eps_1)}\,\pi_1(\eps, \eps_2)$.
For most values of the exogenous parameters this dynamic has four ``locally stable
equilibria'' (sometimes called local Nash equilibria, in continuous games)---strategy pairs at which neither player can increase its own payoff by
a local adjustment (Fig.~\ref{fig:local_br_alg}, bottom row). We interpret these locally stable equilibria as the possible outcomes of an evolutionary, or myopic, process.

Locally stable equilibria fall into three types: both agents staying, both agents exploring, and anti-coordination. At the ``both explore'' outcome $(\eps, \eps)$, the
common exploration rate lies close to the exploration rate at the interior
local maximum of the single-agent reward curve, for the same distance $d$
and reward benefit $R$ (\SI, Fig.~\ref{sup-fig:locally_stable_location}). When
the distant reward is small and far away (high $d$, low $R$), the ``both
explore'' outcome disappears, leaving three locally stable equilibria; in still
more extreme regimes anti-coordination disappears as well, leaving ``both stay'' as the sole evolutionary outcome. Hence the number of behaviors a myopic pair can converge on falls as the distant reward becomes smaller or more distant---``both explore'' is lost first and anti-coordination next---and wherever more than one outcome remains stable, the outcome the agents reach depends not only on the parameters of the environment but also on the exploration rates they begin with.

\subsection*{Competition produces a social dilemma}

Competing agents often converge to socially suboptimal exploration rates. We define the
collective payoff of a strategy pair as the sum of the two players' payoffs,
$\Pi(\eps_1, \eps_2) \coloneqq \pi_1(\eps_1, \eps_2) + \pi_2(\eps_1,
\eps_2)$, and we measure every outcome against the globally optimal
collective payoff,
\[
\Pi^* \coloneqq \max_{(\eps_1, \eps_2) \in [0,1]^2} \Pi(\eps_1, \eps_2),
\]
the largest collective payoff achievable by any strategy pair. A strategy
pair is \emph{socially suboptimal} if its collective payoff falls short of
$\Pi^*$.

A social dilemma arises when the strategy pair that self-interested agents reach is socially suboptimal. Such a dilemma can arise in three nested kinds, each more demanding than the last:

\begin{enumerate}
\item \emph{At least one locally stable equilibrium is socially suboptimal.} Myopic
  agents may converge to a strategy pair whose collective payoff falls short
  of the global optimum $\Pi^*$.
\item \emph{One locally stable equilibrium is collectively worse than another
  locally stable equilibrium.} From some initial conditions, the local dynamics converge to an outcome collectively worse than the one reached from other initial conditions.
\item \emph{Some Nash equilibrium is collectively worse than another
  locally stable equilibrium.} Even agents playing global best responses may remain at a suboptimal Nash equilibrium: neither can improve its payoff alone, yet another dynamically
  achievable outcome yields a higher collective payoff. We call a dilemma of this third kind a \emph{strict} social dilemma; we focus on strict social dilemmas for the rest of our study.
\end{enumerate}

Every Nash equilibrium is also locally stable, so each strict dilemma is also a dilemma of the second kind, and each dilemma of the second kind is a dilemma of the first kind.

\begin{figure*}[tb]
\centering
\includegraphics[]{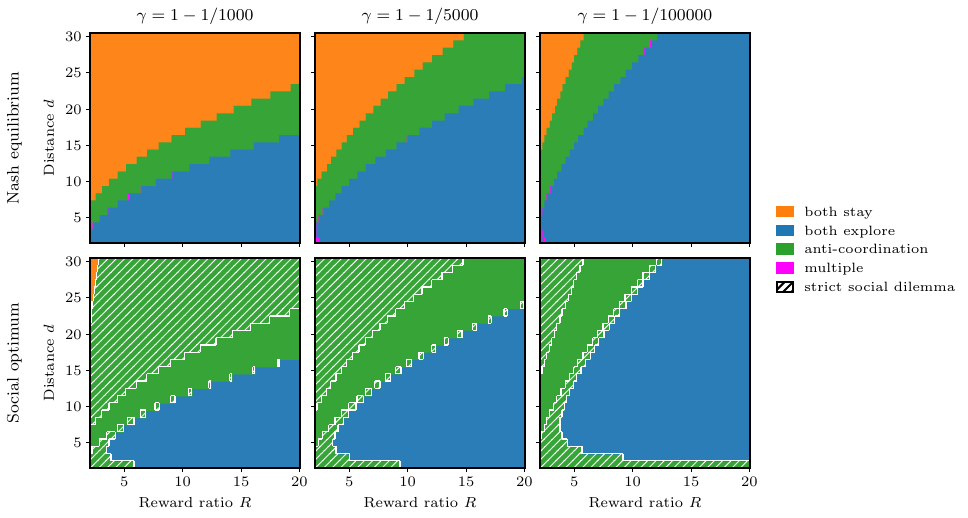}
\caption{\textbf{Strict social dilemmas occur on both sides of the anti-coordination regime, and not within it.} Top: the type of the Nash equilibria as a function of the reward ratio $R$ and the distance $d$, for three discount factors $\gamma$. Orange marks both agents staying, blue both agents exploring, green anti-coordination, and magenta the parameters at which both a both-explore and an anti-coordination Nash equilibrium exist. Bottom: the socially optimal locally stable equilibrium over the same parameters, in the same colors, with white hatching marking the parameters at which a strict social dilemma occurs---that is, where some Nash equilibrium has a lower collective payoff than another locally stable equilibrium. At large distance and small reward ratio the Nash equilibria are of the both-stay type, because exploration is too costly; as the distance decreases and the reward ratio increases, the Nash equilibria are first of the anti-coordination type and then of the both-explore type. Parameters: $R=2.1$--$20$, $d=2$--$30$, $\gamma\in\{1-1/1000,\,1-1/5000,\,1-1/100000\}$ from left to right.}
\label{fig:nash_+_social_optima}
\end{figure*}

Social dilemmas of the first kind are ubiquitous. For any distance $d\geq2$, any positive reward values and any discount factor $\gamma$, both agents staying is a locally stable equilibrium, and every anti-coordination pair has a higher collective payoff (\SI, section~\ref{proof:type-1-dilemma}).
In many cases, the locally stable equilibrium at which both agents stay is far worse than the global social optimum (\SI, Fig.~\ref{sup-fig:def1-def2-social-fitness}).
Averaged over initial pairs of exploration rates, the collective payoff reached under the local best-response dynamics---what a randomly initialized pair can expect---lies near the global optimum at short distances, except at the shortest distance, $d=1$, when the reward ratio is small, and falls furthest below it at large reward ratios and intermediate distances (\SI, Fig.~\ref{sup-fig:social_opt_weighted}).

Dilemmas of the second kind are also nearly universal. The only exceptions are parameters at which every locally stable equilibrium has the same collective payoff (\SI, Table~\ref{tab:social-dilemma-rates}). An anti-coordination pair always has a higher collective payoff than ``both stay,'' so when both staying and an anti-coordination pair are locally stable, there is guaranteed to be a dilemma of the second kind. The collective payoff at an anti-coordination pair $(\eps,0)$ equals exactly the both-stay payoff $1/(1-\gamma)$ plus the reward the exploring agent collects from the distant site: the distant reward $R$ times the expected discounted time the explorer spends there, $\hat\beta(\eps)$, which is positive whenever the explorer reaches site $d$ with positive probability (\SI, section~\ref{sec:coocc}). Among anti-coordination pairs the collective payoff varies with the explorer's rate only through the time that agent spends at the distant reward, so the socially optimal rate of exploration is set by the distance $d$ and the foraging horizon alone: raising the reward ratio $R$ changes which behaviors are equilibria without changing that rate (\SI, section~\ref{sec:coocc}).

Strict dilemmas occur only for a subset of the exogenous parameters, and in every case the suboptimal Nash equilibrium is of the both-stay or the both-explore type, never anti-coordination: in the plane of reward ratio $R$ and distance $d$, strict dilemmas occur on the two sides of the region in which the Nash equilibria are of the anti-coordination type, and not within it (Fig.~\ref{fig:nash_+_social_optima}).
With reward benefit $R = 2.1$ and distance $d = 4$, for example, the ``both explore'' Nash equilibrium achieves a collective payoff only 84\% of that at the best locally stable equilibrium, and yet it is reached from 83\% of initial conditions (Fig.~\ref{fig:local_br_alg}, first column).

At large distances $d$ and small reward ratios $R$, Nash equilibria are of the both-stay type, because the gain from reaching the distant reward does not exceed the opportunity cost of travel. Both agents staying is a Nash equilibrium exactly when the reward ratio is at most half the critical value $R^*(d)$ at which a lone agent begins to explore, $R\leq R^*(d)/2$, because an agent facing a competitor that stays faces the single-agent problem with the reward ratio doubled; the boundary of the both-stay region in Fig.~\ref{fig:nash_+_social_optima} therefore lies at half the reward ratio of the boundary between exploiting and exploring in Fig.~\ref{fig:single-agent-optimal}C (\SI, section~\ref{sec:bothstay}). In part of this regime the anti-coordination pairs, which always have a higher collective payoff than ``both stay,'' remain locally stable, producing a strict dilemma. ``Both stay'' is the socially optimal locally stable equilibrium only where anti-coordination is not itself locally stable; at those parameters no locally stable equilibrium is collectively better than the Nash equilibrium at which both agents stay, and no strict dilemma occurs (Fig.~\ref{fig:nash_+_social_optima}, bottom left).

\begin{table}[tbhp]
\centering
\caption{Nash and social optimality among locally stable equilibria}\label{tab:contingency}
{\footnotesize
\begin{tabular}{@{}llrr@{}}
\toprule
Locally stable & & Socially & Not socially \\
equilibrium & & optimal & optimal \\
\midrule
Both explore, $(\eps,\eps)$ & Nash & 53.7\% & 5.7\% \\
 & Not Nash & 0.0\% & 40.6\% \\
\addlinespace
Anti-coordination, $(\eps,0)$ & Nash & 20.2\% & 0.0\% \\
 & Not Nash & 29.9\% & 49.9\% \\
\bottomrule
\end{tabular}}

\addtabletext{Cross-tabulation of social optimality against whether a strategy pair is a Nash equilibrium, among locally stable equilibria at which both agents explore ($(\eps,\eps)$) and at which the agents anti-coordinate ($(\eps,0)$ and $(0,\eps)$). A locally stable equilibrium is counted as socially optimal when its collective payoff is the largest among the locally stable equilibria at the same parameters. Collected from a sweep of the distance $d=1$--$30$, the reward ratio $R=2.1$--$20$, and three discount factors $\gamma\in\{1-1/1000,\,1-1/5000,\,1-1/100000\}$.}
\end{table}

At intermediate distances $d$ and reward ratios $R$, the Nash equilibria are of the anti-coordination type. Across our parameter range, whenever anti-coordination is a Nash equilibrium it is also the socially optimal locally stable equilibrium (Table~\ref{tab:contingency}), so anti-coordination is never the suboptimal Nash equilibrium of a strict dilemma. At smaller distances and larger reward ratios the Nash equilibria are of the both-explore type. In the narrow interval of reward ratios at the border between these two regions, Nash equilibria of both types are present at once, and a strict dilemma occurs wherever the both-explore equilibrium is collectively worse than anti-coordination.

Strict dilemmas also occur where both agents explore at the Nash equilibrium, at reward ratios just above those at which anti-coordination is the Nash equilibrium (Fig.~\ref{fig:nash_+_social_optima}). There the pair would collect more if one agent stayed at the near reward while the other explored, but neither agent gains by being the one to stop exploring. These dilemmas occur at larger reward ratios $R$ for larger distances $d$. Anti-coordination remains locally stable through most of this range, but it is not a Nash equilibrium. We write $\eps_N$ for the common rate of the both-explore equilibrium and $\eps_{AC}$ for the rate of the exploring agent at anti-coordination. Where these dilemmas occur the reward ratio is high enough that an agent facing a competitor at the symmetric rate earns more by exploring at that rate than by staying, $\pi(\eps_N,\eps_N) > \pi(0,\eps_N)$, but not so high that two explorers collectively out-earn one explorer and one stayer, $2\pi(\eps_N,\eps_N) < \pi(0,\eps_{AC}) + \pi(\eps_{AC},0)$.

The strict dilemmas on the two sides of the anti-coordination region share one explanation: a division of labor, in which one agent stays at the near reward while the other explores, achieves a higher collective payoff than either suboptimal Nash equilibrium. Under that division the near reward is collected on every step---by the agent that stays alone, or split between the two when the explorer returns to site $s=0$, so that the pair receives the whole of it either way---and the distant reward is never contested. Where the gain from reaching the distant reward does not exceed the opportunity cost of travel, both agents stay instead. Where the reward ratio is high enough that exploring pays against a competitor at the symmetric rate, both agents explore instead: each spends time in transit collecting nothing, the near reward goes uncollected much of the time, and whenever both reach the distant reward they must split it. Exploring is then individually worthwhile---each agent earns more by exploring at that rate than it would by staying---but it lowers the collective payoff. Only when the distant reward is large enough does the distant reward gained by a second explorer exceed the near reward forgone when neither agent remains at site $s=0$, and both agents exploring becomes socially optimal as well as a Nash equilibrium (Table~\ref{tab:contingency}, \SI, Fig.~\ref{sup-fig:global-social-optimum}).

\subsection*{Strict dilemmas usually involve a conflict of interest}

Most strict social dilemmas are conflicts of interest between the two agents: no alternative locally stable equilibrium makes both agents better off. Reaching a collectively better outcome requires one agent to change its behavior, and that agent receives a lower payoff: if both agents stay at the Nash equilibrium, it must start exploring; if both explore, it must stop. A mechanism that relies on each agent's self-interest therefore cannot reach the better outcome in these dilemmas.

We distinguish \emph{collective} from \emph{Pareto} dilemmas, separately for each of the three kinds of social dilemmas we have defined. A collective dilemma requires only that the collective payoff of the suboptimal pair $(\eps_1,\eps_2)$ fall short of that of an alternative pair $(\tilde\eps_1,\tilde\eps_2)$,
\[
\pi(\eps_1,\eps_2)+\pi(\eps_2,\eps_1) < \pi(\tilde\eps_1,\tilde\eps_2)+\pi(\tilde\eps_2,\tilde\eps_1).
\]
A Pareto dilemma requires in addition that each agent individually receive a higher payoff at the alternative strategy pair,
\[
\pi(\eps_1,\eps_2) < \pi(\tilde\eps_1,\tilde\eps_2) \quad\text{and}\quad \pi(\eps_2,\eps_1) < \pi(\tilde\eps_2,\tilde\eps_1),
\]
so that the alternative strategy pair Pareto-dominates the suboptimal one. Every Pareto dilemma is therefore also a collective dilemma. Ties are excluded: the strategy space is continuous, so exact equalities are non-generic. The game of Fig.~\ref{fig:payoff-mat} is a strict Pareto dilemma: both agents receive more at each anti-coordination Nash equilibrium than at the Nash equilibrium at which both explore.

Pareto dilemmas of the first and second kinds are nearly universal, but strict Pareto dilemmas are rare (\SI, Table~\ref{tab:social-dilemma-rates}). Strict Pareto dilemmas resemble a Prisoner's Dilemma in that each agent individually would receive a higher payoff at an alternative, dynamically achievable outcome. Such dilemmas can occur when the Nash equilibria are of the both-stay or the both-explore type, and in either case they lie on the border with anti-coordination in the plane of reward ratio $R$ and distance $d$.

The intuition for the case in which both agents explore at the equilibrium is that the agent that must stop exploring gives up the distant reward entirely. At the symmetric equilibrium both agents explore, and each collects some share of what the distant site provides. Reaching anti-coordination requires one of them to lower its exploration rate to zero, after which it never reaches the distant site at all and its payoff consists entirely of the near reward, which it now shares only on the time steps when its competitor happens to return there. That agent gains the share of the near reward it used to give up and loses everything it used to collect from the distant site. The change to anti-coordination raises that agent's payoff only when, at the symmetric equilibrium, exploring earns it little more than staying would, which is why the strict dilemmas that admit a Pareto improvement lie along the border with anti-coordination rather than in the interior of the region where both agents explore.

\subsection*{Signaling can remove suboptimal equilibria} \label{sec:A-switch}

\begin{figure*}[tb]
\centering
\includegraphics[]{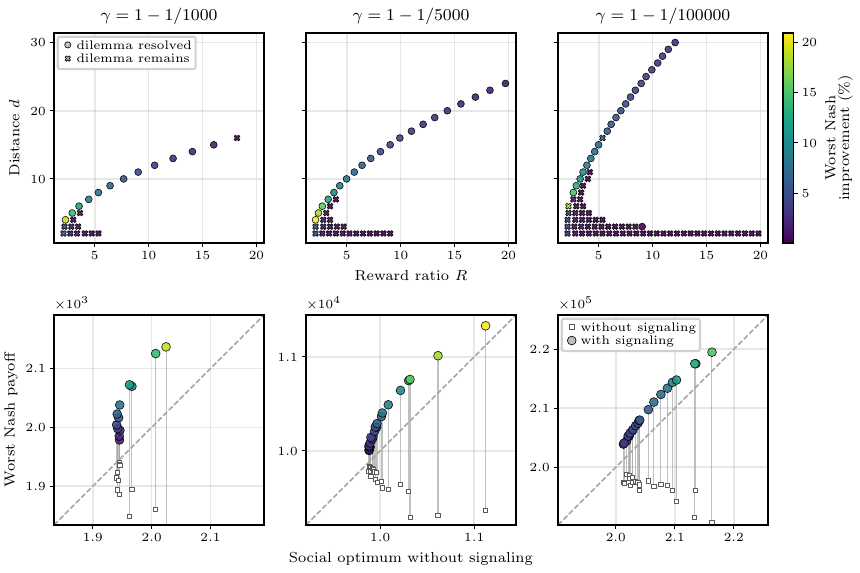}
\caption{\textbf{Signaling often removes suboptimal Nash equilibria where both agents explore, with gains of up to 20.9\% in collective payoff.} Top: parameter combinations at which a strict social dilemma occurs with both agents exploring at the worst Nash equilibrium, that is, the Nash equilibrium with the lowest collective payoff, placed by reward ratio $R$ and distance $d$, for three foraging horizons. Filled circles mark dilemmas that signaling removes and crosses those that remain; color gives the percentage improvement in the collective payoff at the worst Nash equilibrium. Bottom: for each of those parameter combinations, the collective payoff at the worst Nash equilibrium before signaling (open square) and after it (filled circle), joined by a vertical line, plotted against the social optimum without signaling; the dashed diagonal marks a worst Nash equilibrium that attains the social optimum without signaling. The improvement is largest where the collective payoff at the worst Nash equilibrium falls furthest below the social optimum without signaling. Parameters: $R=2.1$--$20$, $d=2$--$30$, $\gamma\in\{1-1/1000,\,1-1/5000,\,1-1/100000\}$ from left to right. The grid, the reward ratios shown, and the omitted parameter combinations are described in \SI, section~\ref{sec:signaling-implementation}.}
\label{fig:signaling-social-dilemma}
\end{figure*}

Can communication of the simplest kind remove the suboptimal Nash equilibria of the preceding sections, including those of strict Pareto dilemmas? In the model so far, each agent uses a single exploration rate from the first step to the last, and the two agents interact only by splitting rewards. Real organisms respond to cues that a competitor has located food, and even a hardwired rule can encode such a response. We therefore consider a modification of the model in which each agent can change its exploration rate $\eps$ upon receiving a signal that the other agent has found the distant reward at site $s=d$.

Suppose one agent, when reaching site $s=d$, sends out a signal to the other agent that it has found the distant reward. The signal may be sent actively, as a sound or a pheromone, or may represent the other agent's passive observation of the discovery. When the first agent sends this signal, the other agent deterministically changes its exploration rate. (In the event they both reach the reward at the same time, we assume they both keep their original exploration rate.) A strategy in this signaling model is therefore an ordered pair $(\eps_i, \eps_i')$, where $\eps_i$ is agent $i$'s exploration rate before any signal and $\eps_i'$ represents its updated exploration rate after receiving a signal from agent $j$. A pair of signaling strategies is locally stable if neither agent can increase its payoff by changing one or both of its two exploration rates by one grid step. This model is implemented as a block Markov chain with four phases: one for the initial stage, two terminal phases when either agent reaches $d$ alone, and another terminal phase if they both reach $d$ at the same time. A full construction is given in \SI, section~\ref{sup-sec:switch_at_A}. The signal does not communicate the size of the distant reward $R$, so an agent receiving it cannot infer that the distant reward exceeds the near one.

The intuition is that the signal is received exactly when a division of labor becomes worth adopting. Before either agent has found the distant reward, the two agents face the same problem, and both explore. Once one of them reaches site $d$ it becomes biased toward that site and spends most of its time there, so from that moment any further exploration by the other agent would move it to a site whose reward it must share, while the near reward is available to it alone. For the agent that receives the signal the near reward is therefore worth more, and the distant reward worth less, than a moment earlier, and lowering its exploration rate becomes the better response. The pair can therefore reach an anti-coordination outcome without either agent having to lower its exploration rate to zero before the distant reward has been found.

Signaling can replace a suboptimal Nash equilibrium with a socially optimal one. For example, without signaling, the game at distance $d = 4$ and reward ratio $R = 2.1$ (Fig.~\ref{fig:payoff-mat}) has four locally stable equilibria and three Nash equilibria, including a suboptimal Nash equilibrium at which both agents explore, which has a large basin of attraction (Fig.~\ref{fig:local_br_alg}) and makes the game a strict Pareto dilemma: each anti-coordination Nash equilibrium gives both agents a higher payoff. With signaling, that both-explore pair is no longer a Nash equilibrium, and the only Nash equilibrium is the pair at which each agent's strategy is $(0.36, 0)$ (Table~\ref{tab:signaling-locally-stable-points}): both agents explore until one reaches the distant reward, after which the other stays at the near reward. The collective payoff at this equilibrium exceeds that of every strategy pair available without signaling. Signaling therefore removes the strict Pareto dilemma of this example, and it raises the collective payoff at the worst Nash equilibrium by $21\%$, from $9357$ to $11362$, on the 51-rate grid of Table~\ref{tab:signaling-locally-stable-points} (it raises the payoff 20.9\% on the finer grid of Figure~\ref{fig:signaling-social-dilemma}). The Nash equilibrium under signaling is not the global optimum of the signaling game: its collective payoff is $97.8\%$ of the largest collective payoff over all pairs of signaling strategies, $11612$, which is attained at a pair that is not a Nash equilibrium.
All locally stable equilibria under signaling besides $((0, 0), (0, 0))$, at which both agents stay, have a higher collective payoff than the both-explore locally stable equilibrium of the baseline model. Signaling therefore does not only remove the suboptimal Nash equilibrium: it replaces it with a Nash equilibrium that is socially optimal, and whose collective payoff exceeds that of every strategy pair available without signaling.

\begin{table}[tbhp]
\centering
\caption{A signal that a competitor has found the distant reward removes the suboptimal Nash equilibrium}\label{tab:signaling-locally-stable-points}
{\footnotesize\setlength{\tabcolsep}{4pt}
\begin{tabular}{@{}lrccc@{}}
\toprule
Strategy pair & Payoff & Nash & Social opt. & Loc. stable \\
\midrule
\multicolumn{5}{@{}l}{\textit{No signaling}, $(\eps_1, \eps_2)$} \\
(0.3, 0.0)* & 11082 & Yes & Yes & Yes \\
(0.28, 0.28) & 9357 & Yes & No & Yes \\
(0.0, 0.0) & 5000 & No & No & Yes \\
\addlinespace
\multicolumn{5}{@{}l}{\textit{Signaling}, $((\eps_1, \eps_1'), (\eps_2, \eps_2'))$} \\
((0.36, 0.0), (0.36, 0.0)) & 11362 & Yes & Yes & Yes \\
((0.3, 0.3), (0.0, 0.0))* & 11082 & No & No & Yes \\
((0.36, 0.22), (0.3, 0.0))* & 10849 & No & No & Yes \\
((0.3, 0.22), (0.3, 0.22)) & 9856 & No & No & Yes \\
((0.3, 0.3), (0.0, 0.22))* & 9477 & No & No & Yes \\
((0.0, 0.0), (0.0, 0.0)) & 5000 & No & No & Yes \\
((0.28, 0.28), (0.28, 0.28)) & 9357 & No & No & No \\
\bottomrule
\end{tabular}}
\addtabletext{Locally stable equilibria under the baseline (non-signaling) model (top) and the signaling model (bottom). With signaling a single Nash equilibrium remains, and it has a higher collective payoff than any locally stable equilibrium without signaling. Payoff is the collective payoff of the pair, the sum of the two agents' expected discounted rewards, and the social optimum is the locally stable equilibrium with the largest collective payoff. A signaling strategy is an ordered pair $(\eps_i,\eps_i')$: the exploration rate agent $i$ uses before any signal, and the rate it adopts after its competitor signals that it has reached the distant reward. An asterisk marks a strategy pair $(a, b)$ whose counterpart with the players' labels exchanged, $(b, a)$, is also a locally stable equilibrium but is not listed. When one agent's initial exploration rate is zero it never reaches the distant site, so its competitor never receives a signal and that competitor's post-signal rate $\eps_j'$ is arbitrary; we set $\eps_j' = \eps_j$ in the table, so that each such row stands for a set of locally stable equilibria. Parameters: $d=4$, $R=2.1$, $\gamma=1-1/5000$.}
\end{table}

Signaling removes many of the strict dilemmas in which both agents explore at the Nash equilibrium (Fig.~\ref{fig:signaling-social-dilemma}, top). Where the Nash equilibria are of the both-stay type, signaling does not remove the dilemma, and cannot: a signal is sent only once an agent has reached the distant site, and in that regime neither agent explores enough to reach it. Wherever signaling removes a strict dilemma, every Nash equilibrium that remains is a socially optimal locally stable equilibrium, so the suboptimal Nash equilibrium is replaced by a socially optimal one. Where signaling does remove the dilemma, the collective payoff at the worst Nash equilibrium rises by as much as $20.9\%$ (\SI, Table~\ref{tab:signaling-strong-vs-weak}). That improvement is smallest at large distance and large reward ratio, but in those regimes the worst Nash equilibrium without signaling already lies close to the social optimum, so the possible gain is small (Fig.~\ref{fig:signaling-social-dilemma}, bottom).

Signaling removes even strict Pareto dilemmas, in which each agent individually, and not only the pair collectively, would receive a higher payoff at another locally stable equilibrium. Among strict dilemmas in which both agents explore at the worst Nash equilibrium, and across the three foraging horizons, signaling removes $45$--$100\%$ of Pareto dilemmas but only $8$--$29\%$ of those that fall short collectively alone; and where signaling does remove a dilemma, the resulting gain in collective payoff at the worst Nash equilibrium is larger for the Pareto cases than for the others (\SI, Table~\ref{tab:signaling-strong-vs-weak}). Hence the dilemmas a signal is most likely to remove are those in which both agents, and not only the pair, gain by moving to the better outcome.

\section*{Discussion}

Organisms must constantly weigh exploiting the resources they know against exploring for better ones, and the balance that serves them best is rarely obvious. A forager that explores too much never remains long enough at any resource to profit from it, whereas one that explores too little stays at a poorer resource while a richer one goes undiscovered. The problem becomes harder still when other foragers share the environment, because the best behavior for each forager then depends on what its competitors do. We have asked how a simple, evolutionarily plausible forager should resolve this tradeoff. Rather than model an algorithm that tunes its exploration over time, which demands a cognitive sophistication many animals lack, we take the exploration rate to be a single fixed parameter, an intrinsic $\eps$-greedy tendency shaped by natural selection to suit an animal's ecology.

When two foragers share a lattice and split rewards, the choice of exploration rate becomes a symmetric two-player game with social dilemmas of three nested kinds. In the mildest and most pervasive kind, some locally stable equilibrium is collectively suboptimal; in the most demanding kind, a strict social dilemma, a Nash equilibrium at which both agents stay or both explore is collectively worse than an anti-coordination outcome, in which one agent explores while the other stays at the near reward. Most strict dilemmas are conflicts of interest: the better outcome would lower the payoff of one agent, the one that would have to start exploring when both stay, or to stop when both explore. Only rarely does the alternative Pareto-dominate the equilibrium, leaving both agents better off, as in a Prisoner's Dilemma. A minimal signal, one agent informing the other that it has reached the distant reward, removes many of the dilemmas in which both agents explore at the equilibrium, most often those in which the alternative Pareto-dominates.

The behaviors the model produces have counterparts in foraging animals. The anti-coordination outcome resembles the division of effort documented when ant colonies exploit two feeders, in which some individuals commit to a known source while others continue to range, and colonies of different species differ in how they divide this effort, the kind of intrinsic variation our fixed exploration rate represents \citep{latty2013-KeepingTrackChanges}. The signal in our model also has natural counterparts: foragers routinely respond to cues that a competitor has located food, whether an actively produced pheromone or the observation of another individual moving toward a resource \citep{giraldeau2000-SocialForagingTheory}. Our results suggest that such communication is valuable not only because it conveys where food lies, but because it can replace an equilibrium that is collectively worse than an achievable alternative with a socially better division of effort.

Among ecological models, the anti-coordination equilibrium resembles the producer--scrounger game, in which some individuals locate resources and others exploit what those individuals find \citep{barnard1981-ProducersScroungersGeneral,giraldeau2000-SocialForagingTheory}; but the agent that stays in our model collects a separate resource rather than its competitor's discovery, and a reward at a site both agents occupy is split. McNamara and Houston \citep{mcnamara1985-OptimalForagingLearning} bring learning into optimal foraging theory for a single forager; our signaling model adds the minimal social element to that setting. Real foragers often search by committing to bouts of directed travel, as in intermittent-search and L\'evy-walk models \citep{benichou2011-IntermittentSearchStrategiesa,viswanathan1999-OptimizingSuccessRandom}, and bumblebees combine reward learning with strong spatial fidelity to familiar patches \citep{biernaskie2009-BumblebeesLearnForagea,crall2018-SpatialFidelityWorkers}; these studies concern the efficiency of search, a separate matter from the tradeoff that $\eps$-greedy represents.

Our results relate to several literatures that have developed largely apart from one another. Within reinforcement learning, most treatments of exploration specify how it should change over time rather than what constant rate is best: regret-minimizing bandit algorithms \citep{lai1985-AsymptoticallyEfficientAdaptive,auer2002-FinitetimeAnalysisMultiarmed}, the Gittins index for discounted bandits \citep{gittins1979-BanditProcessesDynamic}, and the annealing schedules of stochastic optimization, which lower a temperature parameter that controls exploration \citep{kirkpatrick1983-OptimizationSimulatedAnnealing}. The schedules that can be proven sufficient are typically more conservative than what performs well in practice \citep{auer2002-FinitetimeAnalysisMultiarmed,kuleshov2014-AlgorithmsMultiarmedBandit}. Dynamic programming solves the underlying problem exactly \citep{bellman1957-DynamicProgramming,puterman1994-MarkovDecisionProcesses,sutton2018-ReinforcementLearningIntroduction}, but it requires a model of the environment held in advance, including the reward at unvisited sites, which is what exploration would discover, and it is unlikely that animals, or humans deciding in natural settings, perform such a computation. Dynamic programming and heuristics such as optimistic initialization \citep{machado2014-DomainIndependentOptimisticInitialization} also return a policy rather than an exploration rate, and so do not say how often a cognitively simple agent should explore. Empirical comparisons find that $\eps$-greedy performs close to the best available algorithms \citep{vermorel2005-MultiarmedBanditAlgorithms,kuleshov2014-AlgorithmsMultiarmedBandit}, but they tune the exploration rate separately for each problem and do not relate it to the structure of the environment. Our single-agent analysis supplies that relation for one spatial environment: the optimal rate increases with the value of the distant reward, and increases and then falls to zero with the distance to it.

The standard multi-armed bandit problem has a single learner. Its multiplayer extensions, in which several learners compete, typically assume competitive exclusion: two agents that choose the same option collide and gain nothing \citep{boursier-SurveyMultiplayerBandits,rosenski-MultiPlayerBanditsMusical,liu2010-DistributedLearningMultiArmed}. \citet{xu2023-CompetingShareableArms} instead allow rewards to be shared, as we do, and find a Nash equilibrium in which the number of agents at each option is roughly proportional to its value, as in the ideal free distribution of foraging theory \citep{fretwell1969-TerritorialBehaviorOther,parker1986-IdealFreeDistributions}. In those models an agent chooses an option directly; in ours its strategy is an exploration rate and travel is costly, so the agents are not distributed in proportion to resource value, and the Nash equilibria instead fall into three discrete types: both agents exploring, one exploring and one staying, and both staying.

Our assumption of a fixed exploration tendency relates our model to work on how nervous systems control exploration: dopamine signals the reward-prediction error of temporal-difference learning \citep{schultz1997-NeuralSubstratePrediction} and has been proposed to encode the opportunity cost of time \citep{niv2007-TonicDopamineOpportunity}, the cost that travel imposes in our model. How the brain arbitrates between habitual and deliberative control, and how animals weigh the opportunity cost of persistence while foraging, are active questions \citep{daw2005-UncertaintybasedCompetitionPrefrontal,constantino2015-LearningOpportunityCost}. Rather than ask how a single agent should time or direct its exploration, we ask what fixed rate is best.

Our analysis of which outcomes are reached, as distinct from which are optimal, places the model within evolutionary game theory: we ask not only whether a strategy pair is a Nash equilibrium but whether myopic, incremental adjustment converges on it \citep{smith1973-LogicAnimalConflict,hofbauer1998-EvolutionaryGamesPopulation}. Certain forms of reinforcement learning converge, in a suitable limit, to the replicator dynamics \citep{borgers1997-LearningReinforcementReplicator}, so an exploration rate adjusted within a lifetime and one shaped by selection across generations obey related dynamics. Our fixed exploration rate can then be treated as an evolved trait, and the consistent, heritable differences in exploration documented as animal personality \citep{dingemanse2010-RecentModelsAdaptive,wolf2010-ExplanatoryFrameworkAdaptive} as the variation on which selection would act.

We have restricted our analysis to a one-dimensional lattice, whereas animals often forage in two or three dimensions. Because the two agents move independently and interact only by splitting rewards, each agent's payoff equals its solitary payoff reduced by half of the reward it collects when it shares a site with its competitor (\SI, section~\ref{sec:coocc}). The solitary payoff and the shared reward are both determined by how often a lone agent occupies each reward site at each time, so the environment enters the game only through those two single-agent quantities. We conjecture that the same qualitative behavior arises in higher-dimensional lattices, in other graphs, and in environments with clustered rewards, with the exploration rates at which an agent reaches the distant reward rescaled by how long discovery takes. The two-site model of \SI, section~\ref{sec:two-site}, which retains distance only through a hitting probability and a hitting time and reproduces the dilemma (\SI, Fig.~\ref{sup-fig:two_site}), is the simplest such case. An unbiased random walk in three or more dimensions is transient, but our walk is biased toward the highest reward found and the lattice is finite; confirming the conjecture in higher dimensions remains for future study.

A second limitation concerns the search rule. An $\eps$-greedy forager does not commit to a bout of exploration: after a single exploratory step it exploits again with probability $1-\eps$, which tends to return it toward the reward it just left, so that at the short distance examined in \SI, Fig.~\ref{sup-fig:hitting_time_prob}, an agent rarely reaches the distant site before returning to the near one unless its exploration rate is high. \citet{osband-DeepExplorationRandomized} make the same point about $\eps$-greedy in a different setting. Real foragers sometimes commit to bouts of travel, and a rule that commits, such as a L\'evy walk \citep{viswanathan1999-OptimizingSuccessRandom}, may perform better even among rules with a single fixed parameter. We retain $\eps$-greedy because our question concerns the exploration--exploitation tradeoff rather than the efficiency of search, and because it is the simplest rule with which to study that tradeoff.

We have also held the environment fixed. In reality an organism may meet similarly structured environments repeatedly, and an exploration rate shaped by selection would reflect that distribution of environments; nor can an agent know that its environment is static, or that a higher reward exists at all, and that uncertainty is part of what makes a nonzero exploration rate worthwhile. If a patch is exhausted and foraging restarts at a fixed rate $\kappa$, so that both agents return to the near reward with no knowledge of the distant one, the two agents face exactly the same game as in the static model with the discount factor reduced from $\gamma$ to $\gamma(1-\kappa)$, as long as that factor remains in the high-discount regime, so the equilibria and the social dilemma are those of the static model with a shorter foraging horizon (\SI, section~\ref{sec:turnover}). Our comparison across discount factors therefore already spans patches from long-lived to quickly renewed. Environmental change that shifts reward values during foraging, while leaving the agents' positions and value estimates intact, remains to be analyzed.

Finally, we have treated only two competing agents. A population model with heritable exploration rates would show whether the socially suboptimal equilibria we identify are the ones that evolve, and whether signaling can itself be favored by selection. Costly discovery and shared rewards are the rule rather than the exception when animals forage together, so both the collectively suboptimal exploration we identify and the minimal signals that can remove it may be widespread in nature.

\section*{Methods}

The analysis uses three constructions: the transition matrix of a single agent, its extension to two agents by a Kronecker product, and a decomposition that uses the linearity of the payoff in the two reward values, so that a whole grid of rewards costs two linear solves rather than one per value.

\subsection*{The single-agent transition matrix}

Let $s\in \{0, 1, 2, \dots, d\}$ represent the current position of the agent and $B\in \{0,1\}$ a binary variable indicating whether or not the agent has ever visited site $d$. If $S = 2(d+1)$ is the number of possible states $(s, B)$, then $P \in \mathbb{R}^{S\times S}$ is the transition matrix with the following rules, which give the Markov chain of the \emph{Model} section explicitly. We do not assume the high-discount regime here, so that the construction covers every discount factor $\gamma$. The position at which the two rewards have equal discounted value, $x^*$, therefore enters the rules: it solves $\gamma^{x^*} = \gamma^{d-x^*}R$, so that $x^* = (d + \log_\gamma R)/2$, and it need not be a site of the lattice. An agent that has visited site $d$ exploits toward site $0$ from positions $s<x^*$, exploits toward site $d$ from positions $s>x^*$, and moves in either direction with equal probability at $x^*$ itself. In the high-discount regime the condition $\gamma^{d}R>1$ gives $\log_\gamma R < -d$ and hence $x^*<0$, so every site lies beyond $x^*$ and the rules below reduce to the two-phase walk described in the \emph{Model} section.

\begin{center}
\footnotesize\setlength{\tabcolsep}{3pt}
\begin{tabular}{@{}llccc@{}}
\toprule
 & Site & Left, to $s-1$ & Stay & Right, to $s+1$ \\
\midrule
\multirow{2}{*}{$B=0$} & $s=0$ & -- & $1-\eps$ & $\eps$ \\
 & $0<s<d$ & $1-\eps/2$ & $0$ & $\eps/2$ \\
\addlinespace
\multirow{6}{*}{$B=1$} & $s=0<x^*$ & -- & $1-\eps$ & $\eps$ \\
 & $s=0\geq x^*$ & -- & $0$ & $1$ \\
 & $0<s<x^*$ & $1-\eps/2$ & $0$ & $\eps/2$ \\
 & $0<s=x^*$ & $1/2$ & $0$ & $1/2$ \\
 & $\max(0,x^*)<s<d$ & $\eps/2$ & $0$ & $1-\eps/2$ \\
 & $s=d$ & $\eps$ & $1-\eps$ & -- \\
\bottomrule
\end{tabular}
\end{center}
Each row gives the probabilities that an agent in state $(s,B)$ steps to site $s-1$, stays at site $s$, or steps to site $s+1$. The flag $B$ is unchanged except that a step onto site $d$ sets $B=1$, so the state $(d,0)$ never occurs; at distance $d=1$, the step from site $0$ moves the agent directly to state $(1,1)$.

\subsection*{The two-agent transition matrix and reward vectors}

Let $P_i$ be the transition matrix governing the behavior of player $i$. We assume $\gamma$ and $R$ large enough that $\gamma^d R/2 > 1$, that is, either player will always prefer the larger reward after visiting it, even when splitting rewards. The joint transition matrix for both players is the Kronecker product \[K = P_1 \otimes P_2\in \mathbb{R}^{S^2 \times S^2}.\] The joint transition matrix is a Kronecker product because each agent's movement rule depends only on its own state $(s_i,B_i)$ and not on where the other agent is, so the two chains run independently. Only player $i$'s reward vector $\mathbf{r}_i \in \mathbb{R}^{S^2}$ depends on the other agent, and only through whether the two share a site.

Let $s_i$ denote the current site of agent $i$. We write the near reward as $R_0$ and the distant reward as $R_d$, so that $R_0 = r(0) = 1$ and $R_d = r(d) = R$.
Given joint state $z = ((s_1, B_1), (s_2, B_2))$, define the function
\[
    r_i(z) = R_0 \frac{\mathbf{1}\{s_i = 0\}}{1 + \mathbf{1}\{s_j = 0\}} + R_d \frac{\mathbf{1}\{s_i = d\}}{1 + \mathbf{1}\{s_j = d\}},
\]
where $j$ denotes the other agent, so that the two agents split a reward when they occupy the same site, and an agent receives the whole of it when it is the sole occupant. Then construct the reward vector according to the rule $[\mathbf{r}_i]_k = r_i(z_k),$ where $z_k$ is the $k$-th joint state in the ordering (the canonical row-major ordering induced by the Kronecker product).

Then the expected discounted payoff is computed as in the single-agent case:
\[
\pi_i = \mathbf{p}_0(I-\gamma K)^{-1}\mathbf{r}_i.
\]
Because the linear solve is usually the most expensive step, we solve for both agents' payoffs at once:
\[
[\pi_1 \; \pi_2] = \mathbf{p}_0(I-\gamma K)^{-1}[\mathbf{r}_1 \; \mathbf{r}_2].
\]

\subsection*{Linearity of the payoff in the two reward values}\label{sec:basis-trick}

Because the reward vector depends linearly on $R_0$ and $R_d$, we need only use two linear solves to compute the expected discounted reward over a fine grid of reward values. Recall the structure of a reward vector
\[
    [\mathbf{r}]_k = R_0 \, \mathbf{1}\{s = 0\} + R_d \, \mathbf{1}\{s = d\}
\]
in the single-agent case, or
\[
    [\mathbf{r}_i]_k = R_0 \frac{\mathbf{1}\{s_i = 0\}}{1 + \mathbf{1}\{s_j = 0\}} + R_d \frac{\mathbf{1}\{s_i = d\}}{1 + \mathbf{1}\{s_j = d\}}
\]
in the two-player case.
The construction is the same in both cases; we give it for a single agent. Define the vectors $\mathbf{b}^{(0)}$ and $\mathbf{b}^{(d)}$ such that
\begin{align*}
    [\mathbf{b}^{(0)}]_k &= \mathbf{1}\{s = 0\}, \\
    [\mathbf{b}^{(d)}]_k &= \mathbf{1}\{s = d\},
\end{align*}
where the $k$-th value in the ordering corresponds to some state $(s, B)$. Then
\[
    \mathbf{r} = R_0\, \mathbf{b}^{(0)} + R_d \, \mathbf{b}^{(d)}.
\]
The value vector $\mathbf{v}$ then satisfies
\[
    (I - \gamma P)\mathbf{v} = R_0\, \mathbf{b}^{(0)} + R_d \, \mathbf{b}^{(d)}
\]
so that
\[
    \mathbf{v} = R_0\left[(I - \gamma P)^{-1}\mathbf{b}^{(0)}\right] + R_d \left[(I - \gamma P)^{-1} \mathbf{b}^{(d)}\right],
\]
with $\pi = \mathbf{p}_0\mathbf{v}$.
Hence a sweep over many reward values costs two linear solves in total, rather than one for every pair $(R_0,R_d)$: solve for $(I - \gamma P)^{-1}\mathbf{b}^{(0)}$ and $(I - \gamma P)^{-1}\mathbf{b}^{(d)}$ once, then recover the payoff at any reward values by scaling and adding.

\subsection*{Numerical computation of equilibria}

Every payoff we report is computed by solving the linear system above rather than by simulating the agents, so the payoff at any given pair of exploration rates is exact up to floating-point error. The only approximation in our results is the discretization of the strategy space: best responses, Nash equilibria and locally stable equilibria are located on a grid of exploration rates in $[0,1]$, and the local best-response dynamics move by one grid step. The grid spacing therefore sets both the resolution at which we locate an equilibrium and the size of the adjustment against which a locally stable equilibrium must be stable. Under the local best-response dynamics both agents update simultaneously, each moving to whichever of its current rate and the two neighboring rates pays most against the other agent's current rate, and to the lower rate when two pay equally; a strategy pair is locally stable if neither agent's payoff increases with a one-step change, and a trajectory that enters a cycle is assigned the first pair of the cycle it reaches. We classify a strategy pair as both agents exploring if the two rates differ by at most one grid step, as anti-coordination if exactly one rate is zero, and as asymmetric otherwise.

For the single-agent results of Fig.~\ref{fig:single-agent-optimal} we use 201 evenly spaced exploration rates ($\Delta\varepsilon = 0.005$), 200 evenly spaced reward ratios $R = 1.05$--$20$, and the distances $d = 1$--$30$.
We use 1001 rates ($\Delta\varepsilon = 0.001$) for Fig.~\ref{fig:payoff-mat}, 201 ($\Delta\varepsilon = 0.005$) for Fig.~\ref{fig:local_br_alg}, and 51 ($\Delta\varepsilon = 0.02$) for Table~\ref{tab:signaling-locally-stable-points}.
The sweeps behind Figs.~\ref{fig:nash_+_social_optima} and \ref{fig:signaling-social-dilemma} and Table~\ref{tab:contingency} cover the distances $d = 1$--$30$ and 200 evenly spaced reward ratios $R = 2.1$--$20$ at each of the three discount factors, and use 401 evenly spaced rates ($\Delta\varepsilon = 0.0025$) plus the six rates $10^{-8}, 10^{-7}, \dots, 10^{-3}$.

\subsection*{Data and code availability}
Code to reproduce all figures and tables is available on GitHub at \href{https://github.com/james-wolfe/competitive-exploration}{github.com/james-wolfe/\allowbreak competitive-exploration}. The generated data are archived on Zenodo at \href{https://doi.org/10.5281/zenodo.23627789}{doi:10.5281/zenodo.23627789}.

\subsection*{Acknowledgments}
We acknowledge support from the US Army Research Office (award W911NF2410393).

\clearpage
\onecolumn
\newgeometry{margin=1in}
\begin{center}
{\LARGE\sffamily\bfseries SI Appendix}\\[6pt]
{\large Competitive exploration in structured environments}
\end{center}
\vspace{10pt}
\setcounter{section}{0}
\renewcommand{\thesection}{S\arabic{section}}
\setcounter{figure}{0}
\renewcommand{\thefigure}{S\arabic{figure}}
\setcounter{table}{0}
\renewcommand{\thetable}{S\arabic{table}}
\setcounter{equation}{0}
\renewcommand{\theequation}{S\arabic{equation}}

\section{Variable distance to the distant reward}\label{sec:variable-distance-poisson}

In this section we relax the assumption that the distance to the distant reward is fixed. Foraging animals might encounter many different environment structures, and so we consider the possibility that $d$ is not one fixed distance but rather a random variable $d_{V} \sim \mathrm{Poisson}(\lambda) + 1$. In general, the optimal exploration rate when $d_{V} \sim \mathrm{Poisson}(\lambda) + 1$ is close to the optimal exploration rate in the static environment $d = \lambda + 1$, although the explore vs. stay boundary is not quite the same (Fig.~\ref{sup-fig:poisson_opt_eps}). For larger discount factors (longer foraging horizons), there is a region where the optimal exploration rate in a static environment is positive, but zero in a variable environment. The opposite is true for smaller discount factors (shorter foraging horizons): there is a region where the optimal exploration rate in a variable environment is positive, but zero in the static environment. For smaller distances $d$, a variable environment encourages slightly larger optimal exploration rates. Intuitively, an agent might want to explore more when there is a possibility of larger, but still manageable, distances, to ensure it reaches the reward in the most likely landscapes. This correspondence suggests that an exploration rate suited to the mean distance is close to the rate that is optimal when the distance varies, except near the boundary between exploring and not exploring, where the two environments disagree.

\section{Behavior of the threshold reward ratio}
\label{sec:threshold}

Let $\hat{\beta}(\eps)$ be the expected discounted time an agent with exploration rate $\eps$ at the distant site $d$, and $L(\eps)$ the expected discounted time it spends away from the site $0$.
Comparing the payoff of exploration with that of staying put ($\eps = 0$), we have
\[
	\pi(\eps) - \pi(0) = R_d \hat{\beta}(\eps) - R_0 L(\eps).
\]
Exploring at some positive rate therefore becomes worthwhile when the reward ratio $R = R_d/R_0$ exceeds
\[
	R^{*}(d, \gamma) = \inf_{0 < \eps \leq 1} \frac{L(\eps)}{\hat{\beta}(\eps)}.
\]
This boundary can be seen in the main text Fig.~\ref{fig:single-agent-optimal}C. Both $\hat{\beta}$ and $L$ can be computed exactly from the agent's transition matrix.

Fig.~\ref{fig:threshold} shows $R^{*}$ over a wide range of distances and foraging horizons $T = 1/(1-\gamma).$ The threshold $R^*$ increases with the distance, faster than linearly beyond the shortest distances ($R^*/d$ increases with $d$ from $d=4$ at $\gamma = 1-1/1000$ and from $d = 7$ at $\gamma = 1-1/100000$ onward), and decreases as the horizon $T$ grows.

\paragraph{Scaling form.}

To leading order, the behavior of the threshold $R^*$ follows from a trade-off in $\delta = 1-\eps$. Before discovery of the site $d$, the agent steps toward $d$ with probability $\eps/2$ and toward $0$ with probability $1-\eps/2$, giving a ratio $\rho = (1-\delta)/(1+\delta)$. It reaches site $d$ after expected time
\[
    E[\tau] = \frac{\rho^{-d}-1-2d\delta}{2\delta^2} \approx \frac{e^{2d\delta}}{2\delta^2},
\]
and thereafter spends a fraction $f_d$ of its time at site $d$, with
\[
    f_d = \frac{\delta}{1-(1-\delta^2)\rho^d}\approx \delta.
\]
Treating $\tau$ as exponentially distributed, discounting reduces the value of the distant reward by a factor $1/(1+E[\tau]/T)$, while an agent exploring at a rate $\eps$ near 1 forgoes nearly all of the reward at site 0. Exploring therefore becomes worthwhile when
\[
    R\delta \gtrsim 1 + \frac{E[\tau]}{T}.
\]
A larger $\delta$ raises the fraction of time spent at site $d$ but lengthens the search time for site $d$ exponentially. The term $E[\tau]/T$ thus switches from negligible to dominant around $E[\tau] \approx T$, at $\delta \approx \ln(T/d^2)/(2d)$. This yields
\[
    R^*\approx \frac{2d}{\ln(T/d^2)}.
\]
Observe this grows faster than linearly in $d$ and decreases in $T$.

Note the approximations for $E[\tau]$ and $f_d$ assume $\delta \ll 1 \ll d\delta$, so the agent explores at a rate close to 1, but still finds $d$ only rarely.

\section{Proof that there is always a dilemma of the first kind when $\mathbf{d\geq 2}$}\label{proof:type-1-dilemma}

A dilemma of the first kind exists whenever some locally stable equilibrium is collectively worse than a strategy pair the agents could reach instead, so it suffices to exhibit one such point and one such alternative. We take both agents staying as the point and anti-coordination as the alternative. First, the ``both stay'' strategy $(\eps_1, \eps_2) = (0, 0)$ is locally stable; the same argument establishes that $\eps = 0$ is a local maximum in the single-agent case. Second, any anti-coordination strategy has a higher collective payoff than both agents staying.

\subsection{Both stay is locally stable }

Local stability at $(0,0)$ means that a small amount of exploration does not pay. We establish local stability by comparing what an exploring agent forgoes against what it can gain. In particular, we consider a lower bound on the expected opportunity cost, a minimal expected opportunity cost, and we consider an upper bound on the expected rewards of exploration.
An upper bound on the expected rewards of exploration minus the minimal opportunity cost amounts to an upper bound on the difference $\pi(\eps) - \pi(0)$. We will find $\eps>0$ such that this difference is less than zero.

\begin{theorem}
    Let $R_0, R_d>0$ and $\gamma\in (0, 1)$. When $d\geq 2$, the point $(\eps_1, \eps_2) = (0, 0)$ is locally stable. That is, there exists some $\delta > 0$ such that for all $\eps$ satisfying $0 < \eps < \delta$, $\pi(\eps, 0) < \pi(0, 0)$.
\end{theorem}

\begin{proof}
    Note that we need only find one $\delta$ and prove $\pi(\eps, 0) < \pi(0, 0)$ for all $\eps < \delta$, because the game is symmetric.
    Further, note that when player 2's exploration rate is 0, the environment of player 1 is static: there is one reward giving $B_0 = R_0 / 2$ per timestep at one end, and another giving $B_d = R_d$ at the other.
    Thus, we need only prove that small exploration results in payoff losses for one player in an environment with rewards $B_0$ and $B_d$.

    Starting from state $s = 0$, time $t = 0$, in one timestep, exploration incurs an opportunity cost of $\gamma B_0$, i.e. by being in state $s = 1$, the player misses out on reward $\gamma B_0$.
    (Note when $d = 1$, the player reaches the other reward immediately, so there is no opportunity cost.)
    Thus, the \emph{expected} one-step opportunity cost of having exploration rate $\eps$ at state 0 is $\eps \gamma B_0$. An exploring agent's opportunity cost is at least this value.

    We will now consider the benefit of exploration, in particular finding an upper bound.
    Let $\tau:= \inf\{t\geq 0: s_t = d\}$ be the hitting time of state $d$. Then the expected discount rate when arriving at $d$ is $\mathbb{E}[\gamma^\tau].$
    Supposing the player does not deviate from $d$ once arriving, an optimistic assumption, their expected payoff would be
    \[
        \mathbb{E}[\gamma^\tau] \frac{B_d}{1-\gamma}.
    \]
    This expression is an upper bound on the expected benefits of exploration. Then, using our lower bound on the opportunity cost and an upper bound on the benefit of exploration, we can say
    \[
        \pi(\eps) - \pi(0) \leq \mathbb{E}[\gamma^\tau] \frac{B_d}{1-\gamma} - \eps \gamma B_0.
    \]
    By definition,
    \[
        \mathbb{E}[\gamma^\tau] = \sum_{t=0}^\infty \gamma^t \, \mathbb{P}(\tau = t).
    \]
    Since the minimum number of steps it would take to reach $d$ is $d$, we have
    \begin{align}
        \mathbb{E}[\gamma^\tau] &= \sum_{t=d}^\infty \gamma^t \, \mathbb{P}(\tau = t) \\
        &\leq \sum_{t=d}^\infty \gamma^t \, \mathbb{P}(\tau \leq t) \\
        &\leq \sum_{t=d}^\infty \gamma^t \binom{t}{d} \, \eps^d \label{eq:binom-bound} \\
        &= \frac{\gamma^d}{(1-\gamma)^{d+1}} \eps^d. \label{eq:negbin} 
    \end{align}
    Line~\eqref{eq:binom-bound} is derived by observing that the agent would have to make at least $d$ rightward steps over the course of $t$ time steps. There are $\binom{t}{d}$ permutations of when $d$ rightward steps would occur. Any rightward step has probability at most $\eps$, so any collection of $d$ rightward steps has probability at most $\eps^d$. Line~\eqref{eq:negbin} is found via an application of the negative binomial identity.

    Then
    \[
        \pi(\eps) - \pi(0) \leq \frac{\gamma^dB_d}{(1-\gamma)^{d+2}} \eps^d - \eps\gamma B_0
    \]
    Solving for $\eps$ such that the right hand side is less than 0 yields
    \begin{align*}
       \eps &< \left(\frac{B_0}{B_d}\frac{(1-\gamma)^{d+2}}{\gamma^{d-1}}\right)^{1/(d-1)}.
    \end{align*}
    Thus, if we select
    \[
        \delta = \left(\frac{B_0}{B_d}\frac{(1-\gamma)^{d+2}}{\gamma^{d-1}}\right)^{1/(d-1)},
    \]
    then for all $\eps < \delta$, $\pi(\eps) - \pi(0)< 0$.
    \end{proof}

Note that we did not specify anything about $B_0$ and $B_d$ other than that they be positive.

\subsection{Anti-coordination is better than both stay}

Having established that both agents staying is locally stable, we now compare its collective payoff with the collective payoff at an anti-coordination pair. The comparison needs no assumption about the reward values beyond positivity, and it holds at every distance.

\begin{theorem}
    When $\eps > 0$, the group payoff of any anti-coordination strategy pair always exceeds that of both stay. That is, $\Pi(\eps, 0) = \pi(\eps, 0) + \pi(0, \eps) > 2\pi(0, 0) = \Pi(0, 0)$
\end{theorem}

\begin{proof}
    Consider any anti-coordination strategy pair $(\eps,0)$. First, note that since one player's exploration rate is 0, they will always remain at state 0, and thus the full reward $R_0$ (whether it is split by the two agents or not) will be collected by the group every time step. Thus, the group payoff will be at least $R_0/(1-\gamma)$.

    It suffices to observe that the probability of the exploring agent to reach site $d$ in $d$ steps is $\eps (\eps/2)^{d-1}$. Once there, the exploring agent will receive discounted payoff $\gamma^dR_d$ in time step $d$. Thus, the direct path's contribution to expected collective payoff is
    \[
        \gamma^dR_d \, \eps\left(\frac{\eps}{2}\right)^{d-1}.
    \]
    Thus, we can immediately conclude
    \[
        \Pi(\eps, 0)\geq \frac{R_0}{1-\gamma} + \gamma^dR_d \, \eps\left(\frac{\eps}{2}\right)^{d-1} > \frac{R_0}{1-\gamma} = \Pi(0, 0).
    \]
\end{proof}

We do not consider the payoff received by the exploring agent \emph{after} reaching $d$, nor any alternative paths to $d$ of length $>d$, both of which would add to the expected group payoff.

Taking the two theorems together: for $d\geq2$ the pair $(0,0)$ is locally stable, and every anti-coordination pair $(\eps,0)$ with $\eps>0$ yields a strictly greater collective payoff, so the globally optimal collective payoff satisfies $\Pi^*>\Pi(0,0)$. Both agents staying is therefore a locally stable equilibrium that is socially suboptimal, and a dilemma of the first kind exists whenever $d\geq2$.

\section{The two-agent payoff in terms of single-agent occupancies}\label{sec:coocc}

The payoff of an agent playing against a competitor can be written entirely in terms of quantities belonging to a single agent alone. That reduction determines how the environment enters the game.

For a single agent with exploration rate $\eps$, let
\[
\hat\alpha(\eps) = \sum_{t\geq0}\gamma^t\,\mathbb{P}(s_t = 0),
\qquad
\hat\beta(\eps) = \sum_{t\geq0}\gamma^t\,\mathbb{P}(s_t = d)
\]
denote the expected discounted numbers of time steps the agent spends at the near and far reward sites, so that its payoff is $\pi(\eps) = R_0\,\hat\alpha(\eps) + R_d\,\hat\beta(\eps)$.

Because each agent's movement rule is unaffected by the presence of the other (main text, \emph{Materials and Methods}), the two trajectories $s_1(t)$ and $s_2(t)$ are independent. For $b\in\{0,1\}$ we have $1/(1+b) = 1 - b/2$, so the reward function of \emph{Materials and Methods} can be written
\[
r_i(z) = R_0\!\left(\mathbf{1}\{s_i=0\} - \tfrac12\mathbf{1}\{s_i=0\}\mathbf{1}\{s_j=0\}\right)
+ R_d\!\left(\mathbf{1}\{s_i=d\} - \tfrac12\mathbf{1}\{s_i=d\}\mathbf{1}\{s_j=d\}\right).
\]
Taking the discounted expectation from the initial state $((0,0),(0,0))$ and factoring the products of indicators by independence gives
\[
\pi_i(\eps_1,\eps_2) = \pi(\eps_i) - \frac{R_0}{2}\,C_0 - \frac{R_d}{2}\,C_d,
\qquad
C_x = \sum_{t\geq0}\gamma^t\,\mathbb{P}\big(s_1(t)=x\big)\,\mathbb{P}\big(s_2(t)=x\big),
\]
where the discounted co-occupancy of site $x$, $C_x$, weights each time step by the probability that both agents occupy site $x$ at that step. Competition subtracts half the reward accrued over the discounted time the two agents spend together at each reward site, and enters in no other way. Every payoff in the two-agent game is therefore determined by two single-agent quantities---the probability that a lone agent with exploration rate $\eps$ occupies the near site at time $t$, and the probability that it occupies the distant site at time $t$---and by no other feature of the lattice. The discounted occupancies $\hat\alpha$ and $\hat\beta$ do not by themselves determine the game, since the co-occupancies $C_0$ and $C_d$ depend on how a lone agent's occupancy of each site is distributed across time steps.

As a special case, at an anti-coordination pair $(\eps,0)$ the agent that stays remains at site $0$ at all times, so $C_0 = \hat\alpha(\eps)$ and $C_d = 0$, giving
\[
\pi(\eps,0) = \tfrac{R_0}{2}\hat\alpha(\eps) + R_d\,\hat\beta(\eps),
\qquad
\pi(0,\eps) = \tfrac{R_0}{1-\gamma} - \tfrac{R_0}{2}\hat\alpha(\eps),
\qquad
\Pi(\eps,0) = \tfrac{R_0}{1-\gamma} + R_d\,\hat\beta(\eps).
\]

The expression for the collective payoff at anti-coordination has a simple interpretation. The agent that stays never leaves site $0$, so the near reward is collected on every time step: when both agents are there they each take half and the pair still receives the whole of it, and when the explorer is away the stayer receives all of it. The agent that stays also never travels to site $d$, so the explorer never shares what it finds there. Anti-coordination therefore leaves the near reward fully collected (collectively) and the distant reward uncontested, and the collective payoff is the payoff of two agents that both stay, $R_0/(1-\gamma)$, plus the reward the explorer collects from the distant site.

Two consequences follow. First, the collective payoff at any anti-coordination pair with $\eps>0$ strictly exceeds the collective payoff when both agents stay: an agent that explores at any positive rate reaches site $d$ with positive probability, so the discounted time it spends at the distant reward, $\hat\beta(\eps)$, is positive. The gap is exactly $R_d\,\hat\beta(\eps)$, which sharpens the lower bound obtained in Section~\ref{proof:type-1-dilemma}.

Second, the near-reward term $R_0/(1-\gamma)$ does not depend on the exploration rate, so the collective payoff at anti-coordination depends on the explorer's rate only through the discounted time it spends at the distant reward, $\hat\beta(\eps)$. In the high-discount regime the movement rule does not depend on the reward values, so $\hat\beta$ is determined by the exploration rate, the distance $d$, and the discount factor $\gamma$ alone. The exploration rate that maximizes the collective payoff among anti-coordination pairs is therefore the rate that maximizes $\hat\beta$, and it does not depend on how valuable the distant reward is. Changing the distant reward $R_d$ changes the explorer's payoff and the collective payoff, and changes which strategy pairs are equilibria, but it leaves both the payoff of the agent that stays and the socially optimal anti-coordination rate unchanged.

\section{When both agents staying is a Nash equilibrium}\label{sec:bothstay}

Here we derive the range of reward ratios in which both agents staying, $(\eps_1,\eps_2)=(0,0)$, is a Nash equilibrium, and we show that in that range it is the only Nash equilibrium, as stated in the main text (\emph{Nash equilibria}). Both results follow from the occupancies of Section~\ref{sec:coocc}. We work in the high-discount regime, in which the movement rule, and hence every occupancy below, does not depend on the reward values, and we write $R=R_d/R_0$ for the reward ratio.

For a single agent with exploration rate $\eps$, let
\[
L(\eps)=\frac{1}{1-\gamma}-\hat\alpha(\eps)=\sum_{t\geq0}\gamma^t\,\mathbb{P}(s_t\neq0)
\]
denote the expected discounted number of time steps the agent spends away from the near site. An agent that never explores collects the near reward on every step, $\pi(0)=R_0/(1-\gamma)$, so the gain from exploring at rate $\eps$ is
\[
\pi(\eps)-\pi(0)=R_d\,\hat\beta(\eps)-R_0\,L(\eps).
\]
That is, exploring forgoes the near reward over the discounted time spent away from site $0$ and collects the distant reward over the discounted time spent at site $d$. A lone agent therefore gains by exploring at some positive rate exactly when the reward ratio exceeds
\[
R^*(d,\gamma)=\inf_{0<\eps\leq1}\frac{L(\eps)}{\hat\beta(\eps)},
\]
the critical reward ratio at which the reward-maximizing exploration rate $\eps^*$ becomes positive (main text Fig.~\ref{fig:single-agent-optimal}C and Fig.~\ref{fig:threshold}). The time an agent spends away from the near site includes the time it spends at the distant site, so $L(\eps)\geq\hat\beta(\eps)$ and $R^*\geq1$, with equality at $d=1$, where every step away from site $0$ is a step at site $d$.

\begin{theorem}\label{thm:bothstay-nash}
Both agents staying is a Nash equilibrium if and only if $R\leq R^*(d,\gamma)/2$.
\end{theorem}

\begin{proof}
By Section~\ref{sec:coocc}, an agent that explores at rate $\eps$ against a competitor that stays receives $\pi(\eps,0)=\tfrac{R_0}{2}\hat\alpha(\eps)+R_d\hat\beta(\eps)$, and each agent receives $\pi(0,0)=R_0/[2(1-\gamma)]$ when both stay. Hence
\[
\pi(\eps,0)-\pi(0,0)=R_d\,\hat\beta(\eps)-\frac{R_0}{2}\,L(\eps).
\]
Because $\hat\beta(\eps)>0$ for $\eps>0$, the right-hand side is non-positive for every $\eps\in(0,1]$ exactly when $2R\leq L(\eps)/\hat\beta(\eps)$ for every such $\eps$, that is, when $2R\leq R^*$. The game is symmetric, so the same condition applies to the other agent.
\end{proof}

The factor of one half in the condition $R\leq R^*/2$ has a simple interpretation. Against a competitor that stays, an agent shares the near reward whenever it is there, so staying is worth half as much as it is to a lone agent, while the distant reward remains uncontested. The agent therefore faces the single-agent problem with the reward ratio doubled (main text, \emph{Strategies and payoffs}), and it gains by exploring once $2R$ exceeds the single-agent threshold $R^*$. At $d=1$ the threshold is $R^*=1$, so both agents staying is never a Nash equilibrium in the range $R>2$ that we consider.

The gain from exploring is smaller against a competitor that explores than against one that stays, and this comparison excludes every other Nash equilibrium wherever both agents staying is a strict Nash equilibrium. Let $G(\eps'\mid\eps)=\pi(\eps',\eps)-\pi(0,\eps)$ denote the gain to an agent from exploring at rate $\eps'$ rather than staying, when its competitor explores at rate $\eps$, and let $p_t(\eps)=\mathbb{P}(s_t=0)$ denote the probability that a lone agent with exploration rate $\eps$ occupies the near site at time $t$.

\begin{theorem}\label{thm:bothstay-unique}
For all exploration rates $\eps',\eps\in[0,1]$,
\[
G(\eps'\mid\eps)-G(\eps'\mid0)=-\frac{R_0}{2}\sum_{t\geq0}\gamma^t\,[1-p_t(\eps')]\,[1-p_t(\eps)]-\frac{R_d}{2}\,C_d(\eps',\eps)\;\leq\;0 .
\]
Consequently, if $R<R^*(d,\gamma)/2$, both agents staying is the only Nash equilibrium.
\end{theorem}

\begin{proof}
By the co-occupancy identity of Section~\ref{sec:coocc}, $\pi(\eps',\eps)=\pi(\eps')-\tfrac{R_0}{2}\sum_t\gamma^t p_t(\eps')\,p_t(\eps)-\tfrac{R_d}{2}C_d(\eps',\eps)$, and $\pi(0,\eps)=R_0/(1-\gamma)-\tfrac{R_0}{2}\hat\alpha(\eps)$. A competitor that stays has $p_t(0)=1$ and never reaches site $d$, so setting $\eps=0$ gives $G(\eps'\mid0)$. Subtracting the two gains and using $\sum_t\gamma^t=1/(1-\gamma)$ and $\hat\alpha(\eps)=\sum_t\gamma^t p_t(\eps)$ gives the displayed identity, both of whose terms are non-positive.

Now let $R<R^*/2$. By the proof of Theorem~\ref{thm:bothstay-nash}, $G(\eps'\mid0)<0$ for every $\eps'>0$, and so, by the identity, $G(\eps'\mid\eps)<0$ for every $\eps'>0$ and every $\eps\in[0,1]$. Whatever its competitor's exploration rate, an agent receives strictly more by staying than by exploring at any positive rate, so no strategy pair with a positive exploration rate is a Nash equilibrium.
\end{proof}

The two terms in the identity of Theorem~\ref{thm:bothstay-unique} correspond to the two rewards. An agent that stays collects the near reward alone whenever its competitor is away from site $0$, so the more its competitor explores, the more staying is worth; and an agent that explores must share the distant reward whenever its competitor is also at site $d$. A competitor's exploration therefore never raises the gain from exploring, and where exploring does not pay against a competitor that stays, it pays against no competitor. Together, Theorems~\ref{thm:bothstay-nash} and~\ref{thm:bothstay-unique} locate the region of main text Fig.~\ref{fig:nash_+_social_optima} in which the Nash equilibria are of the both-stay type: up to the resolution of the grid of exploration rates, it is the set of reward ratios below half the single-agent threshold, $R<R^*(d,\gamma)/2$.

We checked the identity of Theorem~\ref{thm:bothstay-unique} against direct solutions of the two-agent chain at 60 randomly chosen parameter sets ($d=1$--$12$, $1-\gamma$ between $10^{-5}$ and $10^{-2}$, $R=2.1$--$20$, both exploration rates uniform on $[0,1]$), with maximum relative error below $10^{-8}$; and Theorem~\ref{thm:bothstay-nash} at nine combinations of distance and discount factor, where a search over exploration rates using the exact two-agent payoffs finds both agents staying to be a Nash equilibrium at $R=0.99\,R^*/2$ and not at $R=1.01\,R^*/2$ in every case.

\section{Environmental turnover is equivalent to a shorter horizon}\label{sec:turnover}

Here we extend our model to a situation in which a patch does not last forever. Letting the environment turn over at a constant rate leaves the behavior in the game unchanged apart from a shorter effective foraging horizon, so the comparison across discount factors in the main text is also a comparison across rates of environmental change.

Suppose that at each step, independently with probability $\kappa\in(0,1)$, the current patch is exhausted: both agents return to the initial state $((0,0),(0,0))$ with no memory of the distant reward, and begin foraging a fresh environment of the same structure. Write $\pi_i^{\gamma}(\eps_1,\eps_2)$ for the payoff in the static model at discount factor $\gamma$, and $V_i^{\kappa,\gamma}(\eps_1,\eps_2)$ for the payoff under turnover. We will show that the payoff under turnover is simply a constant factor times the payoff of the static model, with a modified discount rate.

Let $\sigma$ be the first turnover time, so that $\mathbb{P}(\sigma > t) = (1-\kappa)^t$. Before $\sigma$ the agents follow the static process, and at $\sigma$ the state is the initial state again; conditioning on $\sigma$,
\[
V_i^{\kappa,\gamma} = \mathbb{E}\!\left[\sum_{t<\sigma}\gamma^t\,r_i(Y_t)\right] + \mathbb{E}[\gamma^\sigma]\,V_i^{\kappa,\gamma}.
\]
Because turnover is independent of the walk, the first term equals $\sum_{t\geq0}\gamma^t(1-\kappa)^t\,\mathbb{E}[r_i(Y_t)] = \pi_i^{\gamma(1-\kappa)}$, and $\mathbb{E}[\gamma^\sigma] = \gamma\kappa/(1-\gamma(1-\kappa))$. Solving and using $1-\gamma(1-\kappa)-\gamma\kappa = 1-\gamma$,
\[
V_i^{\kappa,\gamma}(\eps_1,\eps_2) = \frac{1-\gamma(1-\kappa)}{1-\gamma}\;\pi_i^{\gamma(1-\kappa)}(\eps_1,\eps_2).
\]
Thus the payoff under environmental turnover is just a pre-factor times the payoff in a static environment, with a different discount rate. And the prefactor is the same for both agents and does not depend on the strategies $(\eps_1,\eps_2)$. As a result, best responses, Nash equilibria, locally stable equilibria, and all three kinds of social dilemma are identical to those of the static model at the reduced discount factor $\gamma(1-\kappa)$, provided the reduced discount factor also satisfies the high-discount condition, so that the static model at $\gamma(1-\kappa)$ has the same movement rule, equivalent to a shorter foraging horizon $T_{\mathrm{eff}} = 1/(1-\gamma(1-\kappa))$.%

\section{A two-site model without explicit spatial structure} \label{sec:two-site}

We construct a simplification of the model that discards explicit spatial structure and retains only two quantities through which distance acts: the probability that an exploring agent reaches the distant site, and the expected time it takes to get there. At the one parameter combination we examined, the simplification reproduces the four locally stable equilibria, their ranking by collective payoff, and the shape of the basins of attraction, which suggests that the dilemma does not require the lattice itself.

We consider only two sites, indexed $1$ and $2$ with associated rewards $r_1 < r_2$. The agents start at site $1$. The strategy used by the agent is the following:

\begin{itemize}
    \item With probability $1-\eps$, move toward the best reward the agent has seen. (Moving toward the best reward it has seen may mean staying at the current site, or trying to go to the other site, which arises only if the agent has seen the higher reward and is currently at the site with the lower reward.)
    \item With probability $\eps$, explore, i.e. try to go to the other site.
\end{itemize}

However, there is a cost of exploration, $c(\eps)$, which is an increasing function of $\eps$, and a probability of successful exploration, $s(\eps)$, which is also increasing in $\eps$. We use the gambler's ruin hitting time and probability formulas, respectively, for each function. Both functions depend on $\eps$ and $d$ through the expressions given below, so that distance enters the two-site model through them alone. The resulting process is a Markov decision process, which we construct below.

Note that the discounting is different from that of the full model. Here, for simplicity, and to model as a strict Markov chain, we subtract the expected cost of being at a given state every turn. The subtraction does not account for the fact that the expected cost is incurred as a lump sum rather than accrued over the steps it represents: for example, if the expected time cost was $k$, in the original model, a time cost of $k$ starting from time $t$ would be $\sum_{s = t}^{t+k-1} \gamma^s$, whereas here it is just $k\gamma^t$. Moreover, the rewards incurred in the original model after this movement would start from $\gamma^{t+k}$, whereas here they start from $\gamma^{t+1}$. However, since $\gamma$ is close to 1 (meaning e.g. $\sum_{s = t}^{t+k-1} \gamma^s \approx k\gamma^t$), and our purpose is to reproduce the structure of the original model rather than its exact payoffs, this approximation suffices.

\paragraph{State space.}
Each agent has internal state space
\[
\mathcal{S}=\{U,H,L\}.
\]
(The three states stand for ``unexplored,'' ``high reward,'' and ``low reward.'') The induced \emph{location} mapping is
\[
\ell:\mathcal{S}\to\{1,2\},\qquad
\ell(U)=\ell(L)=1,\quad \ell(H)=2.
\]

\paragraph{Actions.}
Let $\mathcal{A}(x)$ denote the set of available actions in state $x$:
\[
\mathcal{A}(U)=\mathcal{A}(H)=\{\textsf{stay},\textsf{try}\},\qquad
\mathcal{A}(L)=\{\textsf{try}\}.
\]

\paragraph{Policy.}
The mixed policy $\pi_\eps(\cdot\mid x)$ is
\[
\pi_\eps(\textsf{try}\mid x)=\eps,\quad
\pi_\eps(\textsf{stay}\mid x)=1-\eps
\quad\text{for }x\in\{U,H\},
\]
and in $L$ the action is forced:
\[
\pi_\eps(\textsf{try}\mid L)=1.
\]

\paragraph{Transition probabilities.}
We aim to model a similar biased random walk, as before, but without the explicit spatial structure. The probability of success given a try to switch from state $X$ to state $Y$ is given by the function $s_{X\to Y}(\eps)$.
Note that given the agent is in state $X$, there is only one other state besides $X$ that the agent can reach in its next step. From $U,$ the only reachable state is $H.$ From $H,$ the only reachable state is $L.$ From $L,$ the only reachable state is $H.$ If $Y$ is the corresponding reachable next state of $X$, then the transition probabilities $P(\cdot\mid x,a)$ are:
\[P(Y\mid X, \textsf{try}) = s_{X\to Y}(\eps), \qquad P(X\mid X, \textsf{try}) = 1-s_{X\to Y}(\eps), \qquad P(X\mid X, \textsf{stay}) = 1.\]
These transition probabilities give a matrix over the three internal states:
\[
\begin{bmatrix}
    1- \eps \cdot s_{U\to H}(\eps) & \eps \cdot s_{U\to H}(\eps) & 0 \\
    0 & 1-\eps\cdot s_{H\to L}(\eps) & \eps\cdot s_{H\to L}(\eps) \\
    0 & s_{L\to H}(\eps) & 1-s_{L\to H}(\eps)
\end{bmatrix}.
\]
We choose gambler's ruin probabilities of success. Before discovery, an agent that has stepped from site $0$ to site $1$ performs a biased random walk that moves left with probability $1-\eps/2$ and right with probability $\eps/2$, and the probability of a successful switch from $U$ to $H$ is the probability that this walk reaches site $d$ before it returns to site $0$. The probability of a successful switch from $H$ to $L$ is the same: it is the probability of reaching site $0$ before site $d$ when starting at site $d-1$, with the movement now biased to the right. Finally, $s_{L\to H}$ is the probability of reaching site $d$ before site $0$ when starting at site $1$ with the movement biased to the right, as it is for an agent that has already seen site $d$. Thus, with simple algebra on the classic formulas, we can write:
\begin{align*}
    s_{U\to H}(\eps) = s_{H\to L}(\eps) &= \begin{cases}
        \frac{1-\left(\frac{2-\eps}{\eps}\right)}{1-\left(\frac{2-\eps}{\eps}\right)^d} & \text{if } \eps\neq1, \\
        \frac{1}{d} & \text{if } \eps = 1,
    \end{cases}
    \\
    s_{L\to H}(\eps) &= \begin{cases}
        \frac{1-\left(\frac{\eps}{2-\eps}\right)}{1-\left(\frac{\eps}{2-\eps}\right)^d} & \text{if } \eps\neq1, \\
        \frac{1}{d} & \text{if } \eps = 1.
    \end{cases}
\end{align*}
\paragraph{Per-step attempt cost.}
The attempt cost, as we formulate it, depends on the action but not on its result:
\[
\mathrm{cost}(x,a)=
\begin{cases}
c_x(\eps), & a=\textsf{try},\\[4pt]
0, & a=\textsf{stay}.
\end{cases}
\]
Equivalently, the \emph{expected} attempt cost per step for states $x \neq L$ under $\pi_\eps$ is
\[
\mathbb{E}[\mathrm{cost}\mid x]=
\eps\,c_x(\eps).\\[4pt]
\]
Since the action \textsf{try} is forced in $L$, the expected attempt cost in $L$ is just $c_L(\eps)$.
The cost functions $c_x$ are dictated by the expected hitting time in the gambler's ruin:
\begin{align*}
    c_U(\eps) = c_H(\eps) &= \begin{cases} \frac{1}{1-\eps} - \frac{d}{1-\eps} \cdot
        \frac{1-\left(\frac{2-\eps}{\eps}\right)}{1-\left(\frac{2-\eps}{\eps}\right)^d} & \text{if } \eps\neq1, \\
        d-1 & \text{if } \eps = 1,
        \end{cases}
        \\
        c_L(\eps) &= \begin{cases} \frac{1}{\eps-1} - \frac{d}{\eps-1}\cdot
        \frac{1-\left(\frac{\eps}{2-\eps}\right)}{1-\left(\frac{\eps}{2-\eps}\right)^d} & \text{if } \eps\neq1, \\
        d-1 & \text{if } \eps = 1.
    \end{cases}
\end{align*}

The two-site model reproduces the qualitative behavior of the full model (Fig.~\ref{sup-fig:two_site}). For a moderate change in the reward and the same implicit distance, it has two socially optimal anti-coordination locally stable equilibria in roughly the same location, and one suboptimal ``both explore'' locally stable equilibrium that most initial conditions reach under the local best-response algorithm; the basins of attraction have a similar shape. The spatially motivated gambler's ruin functions are not required for this correspondence, which suggests that other environments, including higher-dimensional lattices and other graphs, would yield the dilemmas we find here.

\section{Signaling model: reduce exploration when other agent reaches reward}

We give the Markov-chain construction for a model of two-player search with added signaling: an agent lowers its exploration rate once its competitor has reached the distant reward (Main text, \emph{Signaling can remove suboptimal equilibria}).

\subsection{Technical formulation} \label{sup-sec:switch_at_A}

There are four possible phases in this model, one transient and the rest terminal:
\begin{align*}
    \Phi_0 &: \text{ transient, neither agent has reached $d$,} \\
    \Phi_1 &: \text{ agent 1 reached $d$ first,} \\
    \Phi_2 &: \text{ agent 2 reached $d$ first,} \\
    \Phi_3 &: \text{ both reached $d$ simultaneously.}
\end{align*}
Let $\mathcal{S}$ denote the joint space of the two agents, and further define the partition
\[
\mathcal{S} = S_0 \; \sqcup \; S_1 \; \sqcup \; S_2 \; \sqcup \; S_3,\]
where
\begin{align*}
    S_0 &= \{((x_1, B_1), (x_2, B_2)): x_1 \neq d, \; x_2 \neq d\},\\
    S_1 &= \{((x_1, B_1), (x_2, B_2)): x_1 = d, \; x_2 \neq d\}, \\
    S_2 &= \{((x_1, B_1), (x_2, B_2)): x_1 \neq d, \; x_2 = d\}, \\
    S_3 &= \{((x_1, B_1), (x_2, B_2)): x_1 = d, \; x_2 = d\}.
\end{align*}
Let $P(\eps)$ denote the transition matrix of an agent with exploration rate $\eps$. Then each terminal phase can be modeled by the joint transition kernels
\begin{align*}
    K_1 &:= P(\eps_1) \otimes P(\eps_2'), \\
    K_2 &:= P(\eps_1') \otimes P(\eps_2), \\
    K_3 &:= P(\eps_1) \otimes P(\eps_2).
\end{align*}
Note that in phase $\Phi_3$, both agents retain their original exploration rate $\eps_i$. We expect this choice to make little difference to the results, whether both agents retain their original values or change them.

The transient phase will be modeled slightly differently. We construct $K_0 = P(\eps_1) \otimes P(\eps_2)$, i.e. the agents using their original exploration rates. Although this has the same construction as $K_3$, they will be used differently; in particular, we will restrict to transitions from $S_0$ only. To do this, we restrict rows of $K_0$ to $S_0$, and for each $j = 0, 1, 2, 3$ define
\begin{align*}
    K_{\{0, j\}} := (K_0)_{S_0, \, S_j}     %
\end{align*}
(Here we use the subsets and their corresponding indices interchangeably.) Now, we put these together to construct the block matrix $T$ governing the transition probabilities of the system:
\[T = \begin{bmatrix}
    K_{\{0, 0\}} & K_{\{0, 1\}} & K_{\{0, 2\}} & K_{\{0, 3\}} \\
    0 & K_1 & 0 & 0 \\
    0 & 0 & K_2 & 0 \\
    0 & 0 & 0 & K_3
\end{bmatrix}.\]
Each block $K_{\{0,j\}}$ with $j\geq1$ stands in $T$ for its embedding into an $|S_0|\times N$ matrix, with zero columns at the states outside $S_j$, so that each block of $T$ has the dimensions its position requires: a transition out of $S_0$ that enters phase $\Phi_j$ lands in a state of $S_j$.
Let $v = \begin{bmatrix} v_0 & v_1 & v_2 & v_3 \end{bmatrix}^\top,$ and $
r = \begin{bmatrix} r_0 & r_1 & r_2 & r_3
\end{bmatrix}^\top$, where:
\begin{align*}
    v_0, r_0 &\in \mathbb{R}^{|S_0|}, \\
    v_i, r_i &\in \mathbb{R}^{|\mathcal S|} \text{ for } i = 1, 2, 3.
\end{align*}
Note $|\mathcal S| = N$ is the total number of joint states in the model, $(2(d+1))^2$. The value function must solve
\[(I - \gamma T) v = r,\] which yields
\begin{align*}
    (I-\gamma K_{\{0, 0\}})v_0 &= r_0 + \sum_{j=1}^3 \gamma K_{\{0, j\}} (v_j|_{S_j}), \\
    (I - \gamma K_i) v_i &= r_i, \quad i = 1, 2, 3.
\end{align*}
Here $(v_j|_{S_j})$ denotes the restriction of the continuation value to states in $S_j$, corresponding to the value upon first entering phase $\Phi_j.$

\subsection{Implementation}\label{sec:signaling-implementation}

We evaluate the signaling game on the grid of the baseline sweep of Figure~\ref{fig:nash_+_social_optima}, 401 evenly spaced exploration rates plus the six rates $10^{-8}, 10^{-7}, \dots, 10^{-3}$, for every strict dilemma with a worst Nash equilibrium of both agents exploring (we have no reason to believe signaling would solve a dilemma where ``both stay'' is the worst Nash equilibrium), at the distances $d=2$--$30$. There are 128, 252, and 545 parameter combinations for $\gamma = 1-1/1000$, $1-1/5000$, and $1-1/100000$, respectively. We omit $d=1$, because there the collective payoff at the worst Nash equilibrium of every strict dilemma falls short of that at the best locally stable equilibrium by at most $0.022\%$. The six small rates allow the one-step test of local stability to detect whether an agent's payoff increases or decreases as its exploration rate leaves zero. At $d=2$ and $3$ this determines whether a pair at which one agent stays is locally stable: at $d=2$, $\gamma=1-1/100000$ and $R=18.56$, for example, the payoff of the agent that stays falls for deviations of its exploration rate up to $10^{-6}$ and rises for deviations of $10^{-5}$ or more.

For legibility, the top row of main text Fig.~\ref{fig:signaling-social-dilemma} shows every seventh such reward ratio at each distance, 24, 43 and 83 of the 123, 242 and 531 combinations at which a Nash equilibrium remains after signaling, and the bottom row shows the same combinations. The 5, 10 and 14 combinations at which signaling leaves no Nash equilibrium on the grid are omitted from the figure and count as dilemmas that remain in Table~\ref{tab:signaling-strong-vs-weak}. The bottom row uses fixed axis limits, and two removed dilemmas are not shown in full: at $\gamma=1-1/1000$, the baseline worst Nash equilibrium at $d=4$, $R=2.28$ (collective payoff $1792$) lies below the plotted range, and at $\gamma=1-1/100000$, the dilemma at $d=3$, $R=9.03$, whose collective payoffs are near $8.8\times10^5$, lies above it.

\newpage

\section*{Supplementary Figures}

\setcounter{figure}{0}
\renewcommand{\thefigure}{S\arabic{figure}}
\setcounter{table}{0}
\renewcommand{\thetable}{S\arabic{table}}

\begin{figure*}[h]
\centering
\vspace*{\fill}
\includegraphics[width=0.6\linewidth]{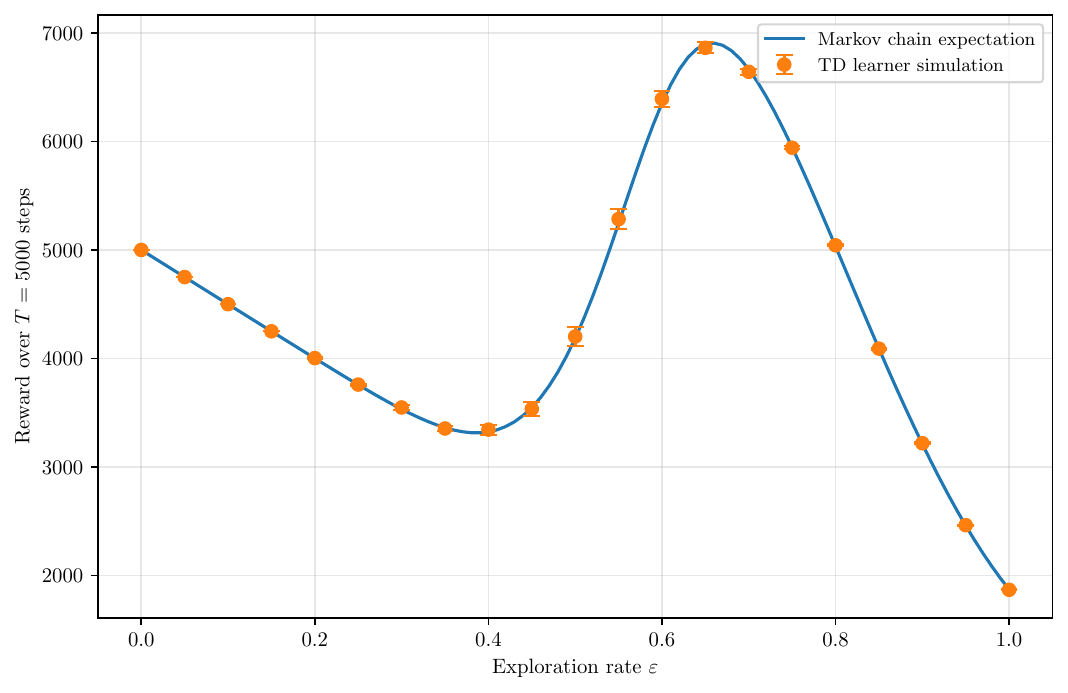}
\caption{\textbf{The Markov chain reproduces the expected reward of the temporal-difference learner.} Expected reward of a single agent over the time steps $t=0,\dots,T$, with $T=5000$, as a function of its exploration rate $\eps$. The curve is the expectation computed from the Markov chain of the main text (\emph{Markov chain formulation}); the points are means over $N=5000$ simulated runs of the temporal-difference learner, with error bars of $\pm 2$ standard errors across runs. The reward is non-monotonic in the exploration rate, with a local maximum at $\eps=0$ and a higher global maximum at an intermediate rate. Parameters: $d=8$, $R=5$, $T=5000$. The Markov chain is evaluated at 101 evenly spaced exploration rates ($\Delta\eps = 0.01$), and the simulations at 21 ($\Delta\eps = 0.05$).}
\label{sup-fig:markov_vs_q}
\end{figure*}

\clearpage

\begin{figure*}[h]
\centering
\vspace*{\fill}
\includegraphics[width=\textwidth]{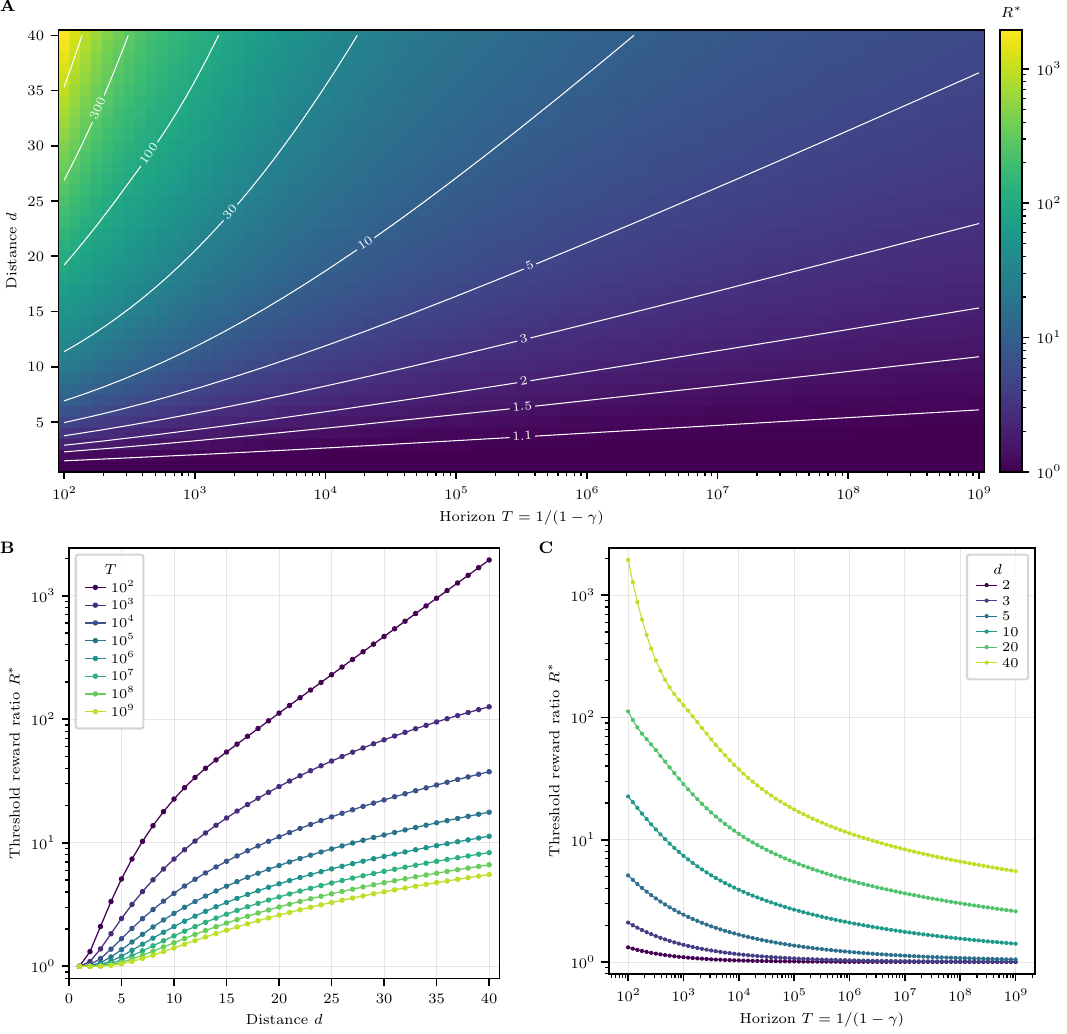}
\caption{\textbf{The reward ratio at which exploration first becomes optimal grows with distance and decreases slowly with the foraging horizon.}
(A) The threshold reward ratio $R^*$ separating exploitation being optimal from exploration being optimal, over the distance $d$ and the horizon $T = 1/(1-\gamma)$; contours are labeled with $R^*$.
(B) $R^*$ against $d$ at fixed time horizons $T$.
(C) $R^*$ against $T$ at fixed distances $d$.
Parameters: $d = 1$--$40$ and $T = 10^2$--$10^9$, log-spaced with 12 values per decade.
$R^*$ is the minimum over $\varepsilon$ of $L(\varepsilon)/\hat\beta(\varepsilon)$, the expected discounted time away from the near site divided by the expected discounted time at the distant site (Section~\ref{sec:threshold}).
$L$ and $\hat{\beta}$ are computed exactly from the agent's Markov chain. The minimum is located on a grid of 410 rates, log-spaced from $10^{-13}$ to $0.01$ and linear from $0.01$ to $0.9995$, and refined with a bounded one-dimensional search, so $R^*$ does not depend on the grid spacing.
The log-spaced rates are needed because, at small $d$ and long horizons, the minimizing rate falls far below $0.01$. At $d = 1$, $R^* = 1$ for every $T$.}
\label{fig:threshold}
\end{figure*}

\clearpage

\begin{figure*}[h]
\centering
\vspace*{\fill}
\includegraphics[width=0.99\linewidth]{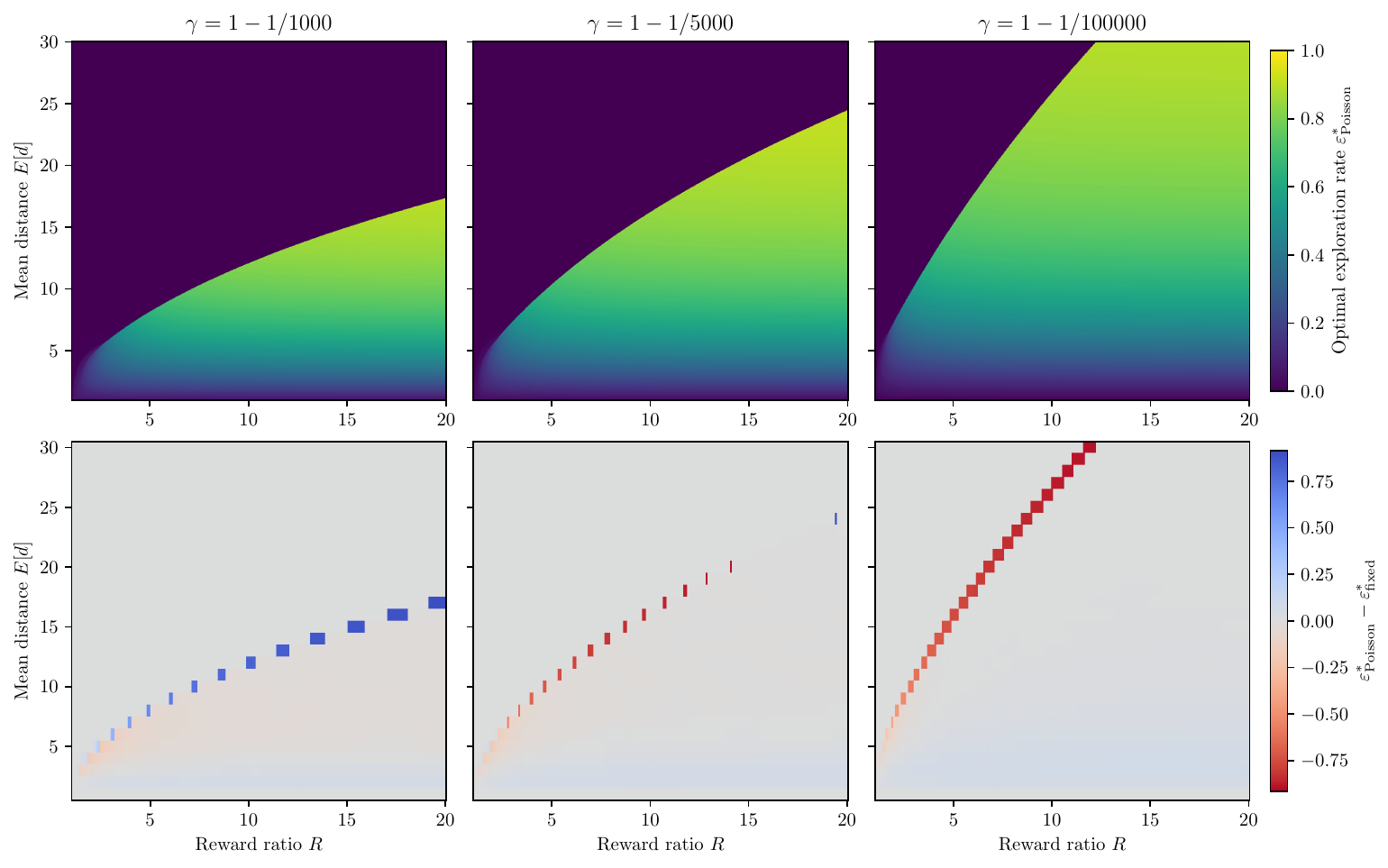}
\caption{\textbf{A random distance to the distant reward leaves the optimal exploration rate nearly unchanged over most of the parameter plane.} Top: the reward-maximizing exploration rate when the distance is random, $d_V \sim \mathrm{Poisson}(\lambda)+1$, over the plane of reward ratio $R$ and mean distance $\mathbb{E}[d_V]=\lambda+1$, for three discount factors $\gamma$, from no exploration in dark to an exploration rate of one in light. Bottom: the difference between that rate and the reward-maximizing rate in a static environment at distance $d=\lambda+1$, $\eps^*_{\mathrm{Poisson}}-\eps^*_{\mathrm{fixed}}$; grey marks agreement, blue a higher rate with the random distance, and red a higher rate with the static distance. The two rates differ only slightly over most of the plane. Along the boundary between exploiting and exploring, the random distance has the higher rate at the shortest foraging horizon and the static distance has the higher rate at the two longer horizons; at the shortest mean distances the random distance has a slightly higher rate at all three horizons. Parameters: $R=1.05$--$20$, $\mathbb{E}[d_V]=1$--$30$, $\gamma\in\{1-1/1000,\,1-1/5000,\,1-1/100000\}$. We maximize over 201 exploration rates ($\Delta\varepsilon = 0.005$), at 1000 evenly spaced reward ratios and 1000 evenly spaced mean distances (top) and at the reward ratios and distances of main text Fig.~\ref{fig:single-agent-optimal}C (bottom).}
\label{sup-fig:poisson_opt_eps}
\end{figure*}

\clearpage

\begin{figure*}[h]
\centering
\vspace*{\fill}
\includegraphics[width=0.99\linewidth]{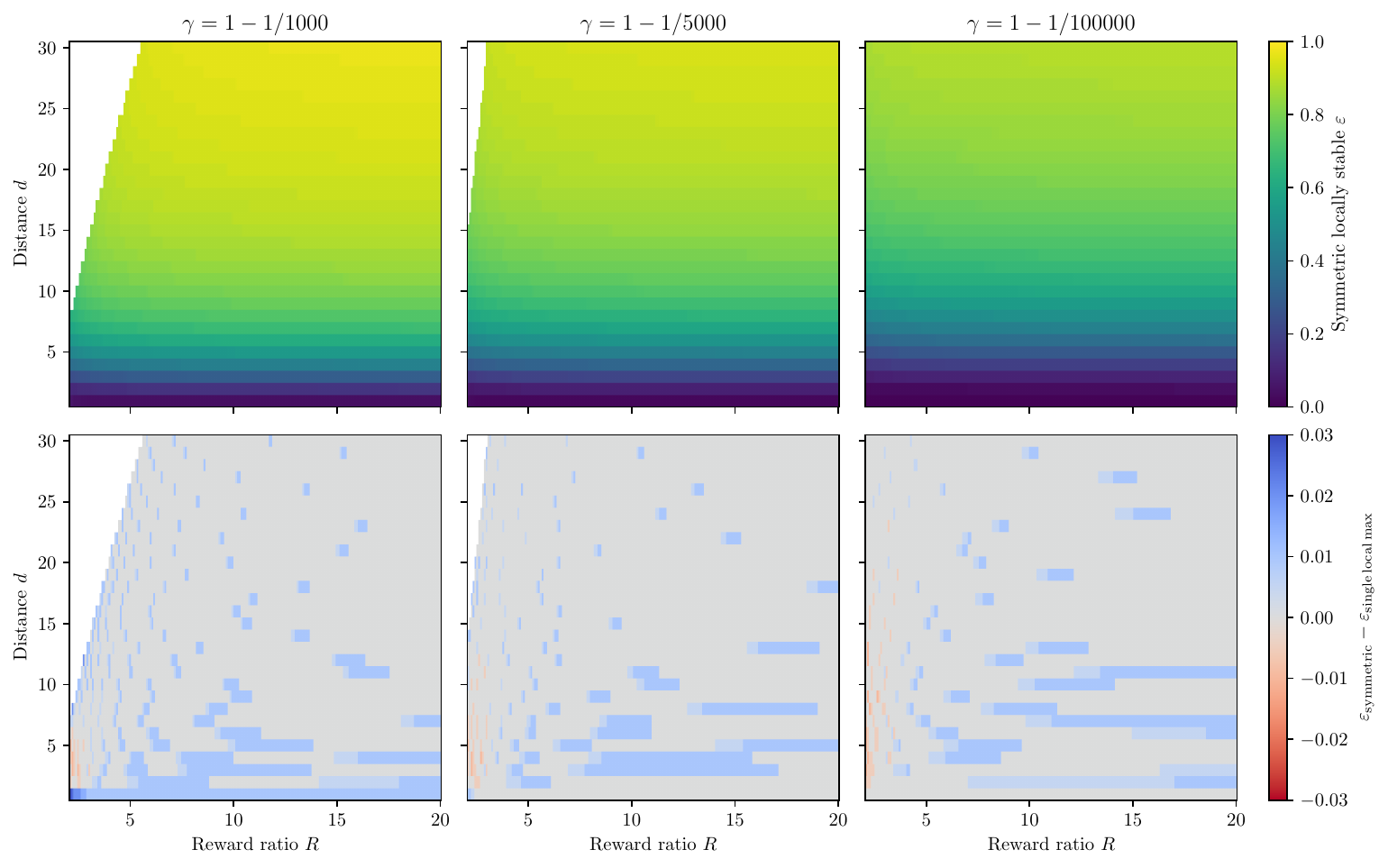}
\caption{\textbf{At the symmetric locally stable equilibrium the two agents explore at close to the rate of the positive local maximum of a lone agent's reward curve.} Top: the common exploration rate at the symmetric locally stable equilibrium $(\eps,\eps)$, over the plane of reward ratio $R$ and distance $d$, for three discount factors $\gamma$; the rate rises with the distance, and white marks the region at large distance and small reward where no such point exists. That region shrinks as the foraging horizon lengthens and is absent at the longest horizon. Bottom: the discrepancy between that rate and the rate at the positive local maximum of the single-agent reward curve, on a scale of $\pm 0.03$, with grey marking agreement. Parameters: $R=2.1$--$20$ (200 evenly spaced values), $d=1$--$30$, $\gamma\in\{1-1/1000,\,1-1/5000,\,1-1/100000\}$. We use 101 evenly spaced exploration rates ($\Delta\eps = 0.01$).}
\label{sup-fig:locally_stable_location}
\end{figure*}

\clearpage

\begin{figure*}[h]
\centering
\vspace*{\fill}
\includegraphics[width=0.99\linewidth]{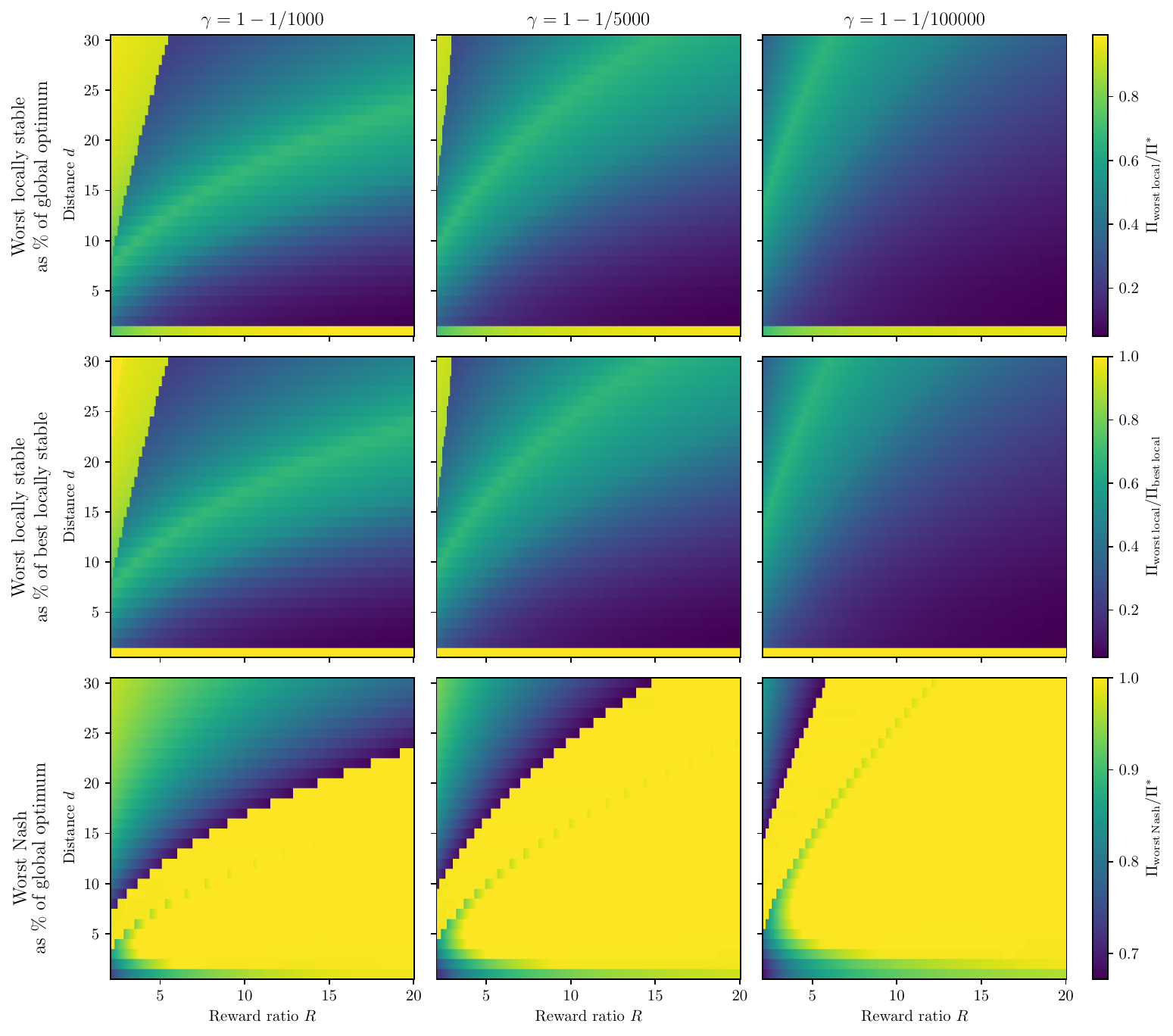}
\caption{\textbf{The worst locally stable equilibrium falls far below both the global optimum and the best locally stable equilibrium across most of the parameter range.} Columns give three discount factors $\gamma$; each row shows one ratio of collective payoffs over the plane of reward ratio $R$ and distance $d$. The three rows use different color scales, dark for a small ratio and yellow for a ratio near one; the bottom row's scale begins near $0.67$, so its darkest cells are far from zero. Top: the collective payoff at the worst locally stable equilibrium as a fraction of the global optimum, $\Pi_{\mathrm{worst\ local}}/\Pi^*$, the comparison that defines a dilemma of the first kind. Middle: the same payoff as a fraction of the best locally stable equilibrium, $\Pi_{\mathrm{worst\ local}}/\Pi_{\mathrm{best\ local}}$, the comparison that defines a dilemma of the second kind. Bottom: the collective payoff at the worst Nash equilibrium as a fraction of the global optimum, $\Pi_{\mathrm{worst\ Nash}}/\Pi^*$; a value of one, in yellow, marks parameters at which even the worst Nash equilibrium attains the global optimum. A strict dilemma is defined by comparison with the best locally stable equilibrium rather than with the global optimum, so this row bounds rather than defines it. At the shortest distance, $d=1$, the worst and the best locally stable equilibria nearly coincide in the top two rows, so the two are nearly equal there. Parameters: $R=2.1$--$20$ (200 evenly spaced values), $d=1$--$30$, $\gamma\in\{1-1/1000,\,1-1/5000,\,1-1/100000\}$. We use 101 evenly spaced exploration rates ($\Delta\eps = 0.01$); for $d \le 5$, where the socially optimal rates are smallest, the rates $10^{-8}, 10^{-7}, \dots, 10^{-3}$ are added.}
\label{sup-fig:def1-def2-social-fitness}
\end{figure*}

\clearpage

\begin{figure*}[h]
\centering
\vspace*{\fill}
\includegraphics[width=0.99\linewidth]{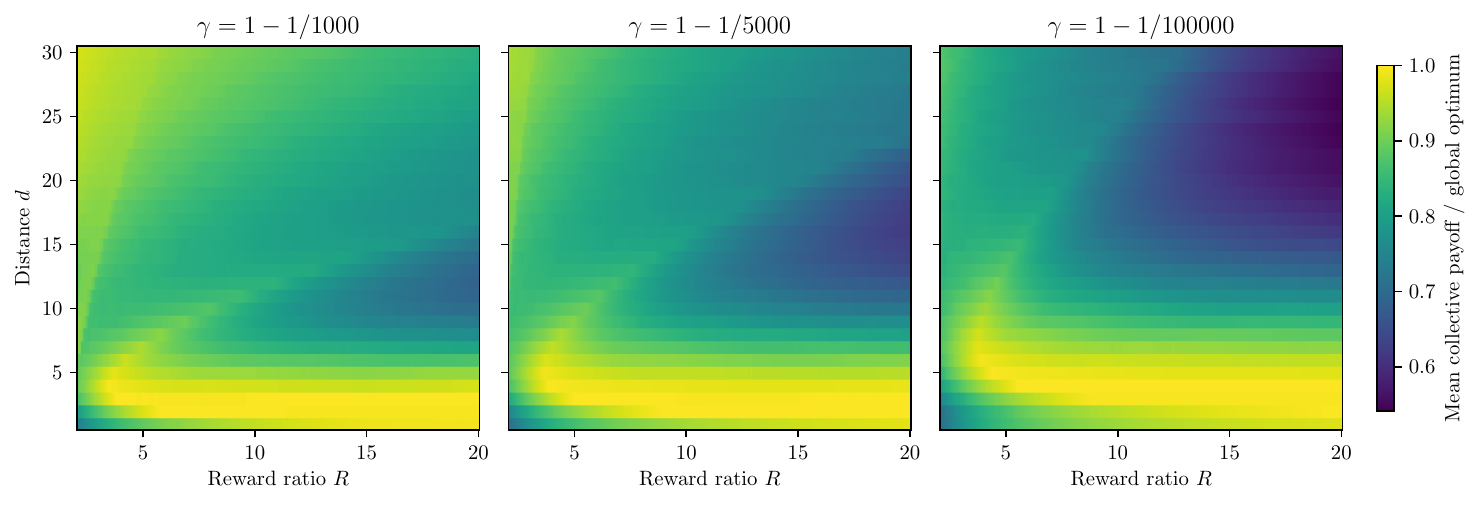}
\caption{\textbf{Myopic agents reach a near-optimal collective payoff at short distances other than the shortest, and the largest shortfall grows and moves to greater distances as the foraging horizon lengthens.} The collective payoff reached under the local best-response dynamics, averaged over initial pairs of exploration rates and expressed as a fraction of the global optimum, over the plane of reward ratio $R$ and distance $d$, for three foraging horizons $T=1000$, $5000$ and $100000$, equivalently discount factors $\gamma=1-1/T$. Dark marks a large shortfall and yellow a collective payoff at the global optimum. The payoff is near the optimum at short distances, falls at the shortest distance, $d=1$, when the reward ratio is small, and reaches its minimum at large reward ratio and an intermediate distance that moves to greater distances as the horizon lengthens. Parameters: $R=2.1$--$20$ (200 evenly spaced values), $d=1$--$30$, with 41 evenly spaced exploration rates ($\Delta\eps = 0.025$). Because both agents update simultaneously, from a few starting pairs, the dynamics alternate between two neighboring strategy pairs instead of settling at one. This occurs at 1052 of the 18000 parameter combinations, from at most 4.4\% of starting pairs at any particular one of them (0.1\% overall). Such trajectories are counted at the first pair of the cycle they reach; excluding them changes the ratio by at most 0.006.}
\label{sup-fig:social_opt_weighted}
\end{figure*}

\clearpage

\begin{figure*}[h]
\centering
\vspace*{\fill}
\includegraphics[width=0.99\linewidth]{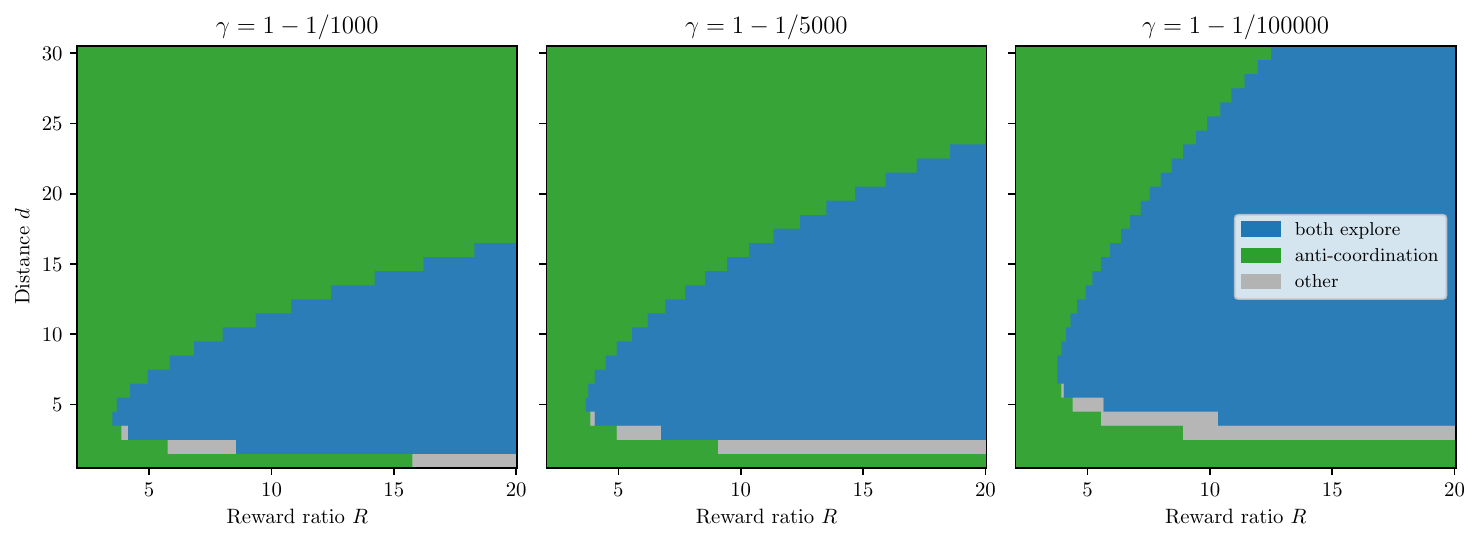}
\caption{\textbf{Anti-coordination maximizes the collective payoff at small reward ratios and at the shortest distances, and both agents exploring maximizes it at larger reward ratios.} The strategy pair maximizing the collective payoff, over the plane of reward ratio $R$ and distance $d$, for three discount factors $\gamma$. Green marks anti-coordination, in which one agent explores while the other stays at the near reward; blue marks both agents exploring; gray marks asymmetric pairs in which both agents explore at different positive rates. Both agents staying never maximizes the collective payoff, by the theorem of Section~\ref{proof:type-1-dilemma}, and so does not appear. The region in which both agents exploring maximizes the collective payoff grows as the foraging horizon lengthens, reaching the largest distances in the sweep at the longest horizon. That region is bounded below by asymmetric pairs and, at the shortest distance, by anti-coordination. Parameters: $R=2.1$--$20$, $d=1$--$30$, $\gamma\in\{1-1/1000,\,1-1/5000,\,1-1/100000\}$, on the grid of main text Fig.~\ref{fig:nash_+_social_optima}: 401 evenly spaced exploration rates ($\Delta\eps = 0.0025$) plus the rates $10^{-8}, 10^{-7}, \dots, 10^{-3}$.}
\label{sup-fig:global-social-optimum}
\end{figure*}

\clearpage

\begin{table}[htbp]
\centering
\setlength{\tabcolsep}{5pt}
\begin{tabular}{l|rr|rr|rr}
 & \multicolumn{2}{c|}{First kind} & \multicolumn{2}{c|}{Second kind} & \multicolumn{2}{c}{Strict} \\
$\gamma$ & Collective (\%) & Pareto (\%) & Collective (\%) & Pareto (\%) & Collective (\%) & Pareto (\%) \\
 \hline
$1-1/1000$ & 100.0 & 100.0 & 99.2 & 92.3 & 48.1 & 0.9 \\
$1-1/5000$ & 100.0 & 100.0 & 100.0 & 98.2 & 26.8 & 0.4 \\
$1-1/10000$ & 100.0 & 100.0 & 100.0 & 99.3 & 21.2 & 0.1 \\
$1-1/50000$ & 100.0 & 100.0 & 100.0 & 100.0 & 16.0 & 0.2 \\
$1-1/100000$ & 100.0 & 100.0 & 100.0 & 100.0 & 15.2 & 0.1 \\
$1-1/500000$ & 100.0 & 100.0 & 100.0 & 100.0 & 14.2 & 0.1 \\
$1-1/1000000$ & 100.0 & 100.0 & 100.0 & 100.0 & 14.2 & 0.1 \\
\end{tabular}
\caption{\textbf{Social dilemmas of the first two kinds occur at nearly every parameter combination, while strict dilemmas admitting a Pareto improvement are rare.} Percentage of parameter combinations at which a dilemma of each kind occurs, for each discount factor $\gamma$. The Collective columns count dilemmas in which the collective payoff falls short of an alternative; the Pareto columns count the subset in which each agent individually also receives a higher payoff at that alternative. Sweep: 200 reward ratios $R=2.1$--$20$, $d=2$--$30$, with 48 exploration rates (41 evenly spaced in $[0, 1]$ plus $10^{-8}, \dots, 10^{-2}$). The distance $d = 1$ is excluded because a dilemma of the first kind is guaranteed only for $d \ge 2$ (Section~\ref{proof:type-1-dilemma}). Strict Pareto dilemmas are a small fraction of strict dilemmas: about 2\% or fewer at every discount factor.}
\label{tab:social-dilemma-rates}
\end{table}

\clearpage

\begin{table}[htbp]
\centering
\begin{tabular}{lrr}
\toprule
& \textbf{Collective only} & \textbf{Pareto} \\
\midrule
\multicolumn{3}{l}{
$\gamma = 1-1/1000$
} \\
Dilemmas solved (\%)             & 28.7 & 83.3 \\
Mean worst-Nash improvement (\%) & 6.0 & 10.2 \\
Largest worst-Nash improvement (\%) & 14.2 & 19.3 \\
\addlinespace

\multicolumn{3}{l}{
$\gamma = 1-1/5000$
} \\
Dilemmas solved (\%)             & 23.2 & 100.0 \\
Mean worst-Nash improvement (\%) & 4.6 & 12.3 \\
Largest worst-Nash improvement (\%) & 16.4 & 20.9 \\
\addlinespace

\multicolumn{3}{l}{
$\gamma = 1-1/100000$
} \\
Dilemmas solved (\%)             & 7.7 & 45.5 \\
Mean worst-Nash improvement (\%) & 4.5 & 8.3 \\
Largest worst-Nash improvement (\%) & 11.0 & 15.1 \\
\bottomrule
\end{tabular}
\caption{\textbf{Signaling removes a larger fraction of Pareto dilemmas than of dilemmas that fall short collectively alone.} Percentage of strict dilemmas resolved by signaling, and the mean and largest improvement in the collective payoff at the worst Nash equilibrium among the dilemmas resolved, for three discount factors $\gamma$. Only strict dilemmas at which both agents exploring is the worst Nash equilibrium are counted, at the distances $d=2$--$30$. At $\gamma=1-1/1000$ there are 122 dilemmas in the Collective-only column and 6 in the Pareto column, of which signaling removes 35 and 5; at $\gamma=1-1/5000$, 246 and 6, of which it removes 57 and 6; and at $\gamma=1-1/100000$, 534 and 11, of which it removes 41 and 5. Signaling is evaluated on a grid of 401 evenly spaced exploration rates plus the rates $10^{-8}, 10^{-7}, \dots, 10^{-3}$ (Section~\ref{sec:signaling-implementation}).}
\label{tab:signaling-strong-vs-weak}
\end{table}

\clearpage

\begin{figure*}[h]
\centering
\vspace*{\fill}
\includegraphics[width=0.97\linewidth]{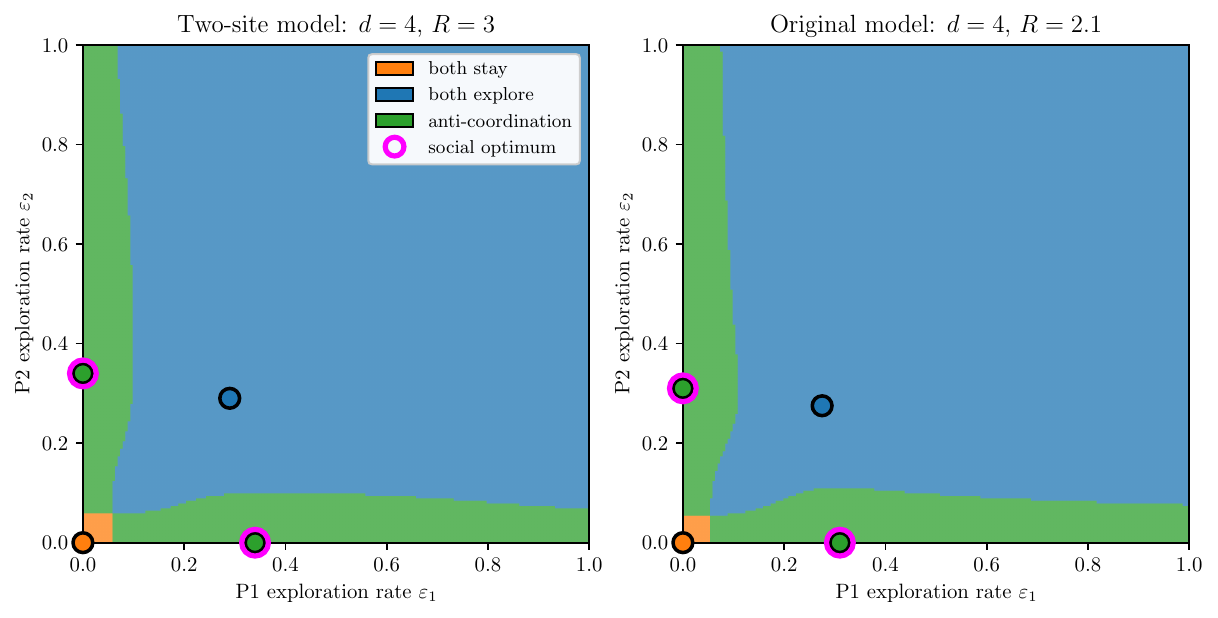}
\caption{\textbf{The two-site model reproduces the equilibrium structure and the social dilemma of the full model.} Locally stable equilibria and their basins of attraction in the unit square of exploration-rate pairs $(\eps_1,\eps_2)$, for the two-site model of Section~\ref{sec:two-site} (left) and the full lattice model (right). Shading gives the outcome reached under the local best-response dynamics from each starting pair: orange for both agents staying, blue for both agents exploring, green for anti-coordination. Circles mark the locally stable equilibria themselves, colored by outcome, and a magenta ring marks the socially optimal ones. Both models have four locally stable equilibria: the two anti-coordination points, which are socially optimal; the symmetric both-explore point, which is not; and both agents staying at the near reward. In both models the basin leading to the suboptimal symmetric point covers most of the square. Parameters: $d=4$, with $R=3$ in the two-site model and $R=2.1$ in the full model. We use 201 evenly spaced exploration rates ($\Delta\eps = 0.005$).}
\label{sup-fig:two_site}
\end{figure*}

\clearpage

\begin{figure*}[h]
\centering
\vspace*{\fill}
\includegraphics[width=0.95\linewidth]{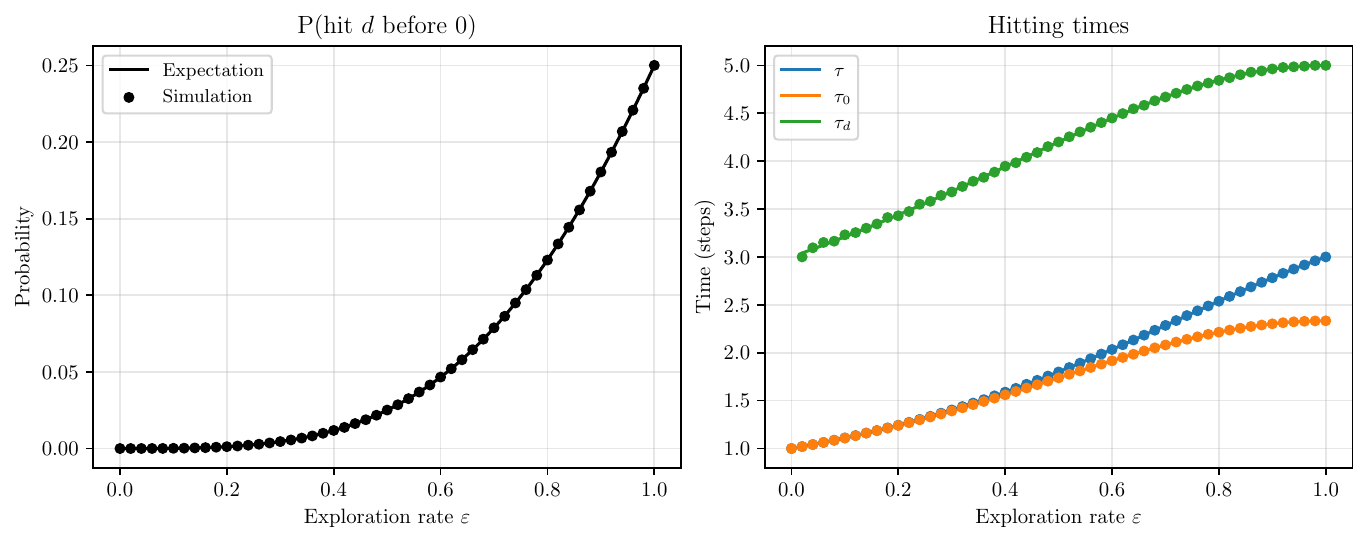}
\caption{\textbf{An $\eps$-greedy walker rarely reaches the distant reward before returning to the near one, and the gambler's-ruin expressions reproduce its hitting statistics.} Left: the probability of reaching site $d$ before returning to site $0$, which stays below $0.01$ until the exploration rate reaches about $0.4$ and rises to $1/d = 0.25$ at $\eps=1$. Curves are the gambler's-ruin expressions of Section~\ref{sec:two-site}; points are simulations of the lattice model. Right: the expected number of steps taken before leaving the interior of the lattice, unconditionally (blue) and conditioned on reaching site $0$ first (orange) or site $d$ first (green), each increasing in the exploration rate $\eps$.  Parameters: $d=4$, at 51 evenly spaced exploration rates ($\Delta\eps = 0.02$); each simulated point is the mean over $10^7$ walks.}
\label{sup-fig:hitting_time_prob}
\end{figure*}

\end{document}